\documentclass[reprint,superscriptaddress,amssymb,amsmath,aps,prx,longbibliography]{revtex4-2}
\usepackage{minitoc}

\usepackage{graphicx}
\usepackage{dcolumn}
\usepackage{bm}
\usepackage{siunitx}
\usepackage{booktabs}
\usepackage[usenames,dvipsnames]{xcolor}
\usepackage{tcolorbox}
\usepackage{tabularx}
\usepackage{array}
\usepackage{colortbl}
\usepackage{braket}
\usepackage{gensymb}
\usepackage{pdfpages}
\usepackage{amsmath}
\usepackage{mathrsfs}
\usepackage{blindtext}
\usepackage{minitoc}
\usepackage{dsfont}
\usepackage{multirow}
\usepackage{xcolor}
\usepackage{soul}

\usepackage{amssymb}

\makeatletter
\patchcmd{\@outputpage@head}{\@ifx{\LS@rot\@undefined}{}{\LS@rot}}{}{}{}
\makeatother

\tcbuselibrary{skins}

\newcommand{\units}[1]{\,\mathrm{#1}}
\tcbset{tab2/.style={colback=gray!5!white,colframe=black!50!black,colbacktitle=Gray!40!white,
coltitle=black,center title}}

\begin{document}

\title{Single-Shot Readout and Weak Measurement of a Tin-Vacancy Qubit in Diamond}

\author{Eric I. Rosenthal}
\thanks{These authors share equal contributions}
\email{ericros@stanford.edu}
\affiliation{E. L. Ginzton Laboratory, Stanford University, Stanford, California 94305, USA}

\author{Souvik Biswas}
\thanks{These authors share equal contributions}
\email{souvik@stanford.edu}
\affiliation{E. L. Ginzton Laboratory, Stanford University, Stanford, California 94305, USA}

\author{Giovanni Scuri}
\affiliation{E. L. Ginzton Laboratory, Stanford University, Stanford, California 94305, USA}

\author{Hope Lee}
\affiliation{E. L. Ginzton Laboratory, Stanford University, Stanford, California 94305, USA}

\author{Abigail J. Stein}
\affiliation{E. L. Ginzton Laboratory, Stanford University, Stanford, California 94305, USA}

\author{Hannah C. Kleidermacher}
\affiliation{E. L. Ginzton Laboratory, Stanford University, Stanford, California 94305, USA}

\author{Jakob Grzesik}
\affiliation{E. L. Ginzton Laboratory, Stanford University, Stanford, California 94305, USA}

\author{Alison E. Rugar}
\thanks{Present address: Apple, Cupertino, CA 95014, USA}
\affiliation{E. L. Ginzton Laboratory, Stanford University, Stanford, California 94305, USA}

\author{Shahriar Aghaeimeibodi}
\thanks{Present address: AWS Center for Quantum Computing, San Francisco, California, USA}
\affiliation{E. L. Ginzton Laboratory, Stanford University, Stanford, California 94305, USA}

\author{Daniel Riedel}
\thanks{Present address: AWS Center for Quantum Networking, Boston, Massachusetts, USA}
\affiliation{E. L. Ginzton Laboratory, Stanford University, Stanford, California 94305, USA}

\author{Michael Titze}
\affiliation{Sandia National Laboratories, Albuquerque, New Mexico 87123, USA}

\author{Edward S. Bielejec}
\affiliation{Sandia National Laboratories, Albuquerque, New Mexico 87123, USA}

\author{Joonhee Choi}
\affiliation{E. L. Ginzton Laboratory, Stanford University, Stanford, California 94305, USA}

\author{Christopher P. Anderson}
\affiliation{E. L. Ginzton Laboratory, Stanford University, Stanford, California 94305, USA}
\affiliation{Department of Materials Science and Engineering, University of Illinois Urbana-Champaign, Urbana, Illinois 61801, USA}

\author{Jelena Vu\v{c}kovi\'{c}}
\affiliation{E. L. Ginzton Laboratory, Stanford University, Stanford, California 94305, USA}

\date{\today}
\begin{abstract}
The negatively charged tin-vacancy center in diamond (SnV$^-$) is an emerging platform for building the next generation of long-distance quantum networks. This is due to the SnV$^-$'s favorable optical and spin properties including bright emission, insensitivity to electronic noise, and long spin coherence times at temperatures above 1 Kelvin. Here, we demonstrate measurement of a single SnV$^-$ electronic spin with a single-shot readout fidelity of 87.4\%, which can be further improved to 98.5\% by conditioning on multiple readouts. In the process, we develop understanding of the relationship between strain, magnetic field, spin readout, and microwave spin control. We show that high-fidelity readout is compatible with rapid microwave spin control, demonstrating a favorable parameter regime for use of the SnV$^-$ center as a high-quality spin-photon interface. Finally, we use weak quantum measurement to study measurement-induced dephasing; this illuminates the fundamental interplay between measurement and decoherence in quantum mechanics, and provides a universal method to characterize the efficiency of color center spin readout. Taken together, these results overcome an important hurdle in the development of SnV$^-$-based quantum technologies, and in the process, develop techniques and understanding broadly applicable to the study of solid-state quantum emitters.
\end{abstract}
\maketitle

\section{Introduction}

Color center qubits have been recognized as an advantageous platform for the realization of quantum technologies, and, in particular quantum networks \cite{kimble:2008}, due to their efficient spin-photon interface, long spin coherence times, and compatibility with nanophotonics \cite{bradac:2019,janitz:2020,wolfowicz:2021}. An outstanding challenge is to scale quantum networks to include more nodes, greater distance between nodes, and further complexity including error corrected registers of qubits within each node \cite{wehner:2018}. 

The path to solving these many challenges is specific to the choice of quantum platform and, in this case, choice of color center. Today, state-of-the-art quantum networks consist of three nodes, where each node is based on a single nitrogen-vacancy center (NV$^-$) in diamond \cite{pompili:2021,hermans:2022}. However, the NV$^-$ is not the optimal qubit for future quantum networks because of its sensitivity to electrical noise and a low probability of emission into its zero-phonon line (ZPL), which reduces entanglement generation rate.

Of countless optically active solid-state atomic defects including rare earth ions \cite{ourari:2023,ruskuc:2024} and defects in silicon carbide \cite{lukin:2020}, group-IV centers in diamond have emerged as promising candidates to advance quantum networks. These qubits have a centrosymmetric structure, which renders a first-order insensitivity to electrical noise, allowing for relative optical stability even within nanophotonic structures \cite{bradac:2019}. Advantageously, these centers also have high quantum efficiency ($80\%$ for the SnV$^-$ \cite{iwasaki:2017}) and strong emission into their ZPL (Debye-Waller factors of approximately $60\%$ \cite{ruf:2021}). This strong, coherent emission promises high rates of entanglement generation compared to quantum networks based on the NV$^-$, which has a Debye-Waller factor of only $3\%$ \cite{faraon:2011}. Among group IV's, the silicon-vacancy center (SiV$^-$) \cite{hepp:2014} is the most developed, with demonstrations of long spin coherence times and single-shot electron spin readout \cite{sukachev:2017}, integration with nanophotonic cavities \cite{sipahigil:2016,zhang:2018,sun:2018,nguyen:2019prl,nguyen:2019,bhaskar:2020,stas:2022}, and demonstration of a two-node quantum network \cite{knaut:2024}.

However, because the SiV$^-$ has the smallest spin-orbit-induced ground state splitting (50 GHz) of all group-IV centers, it is also the most naturally susceptible to decoherence due to thermal excitation out of the spin subspace. State-of-the-art SiV$^-$-based experiments, thus, use highly strained SiV$^-$'s to increase this splitting and reduce susceptibility to drive-induced heating, and, furthermore, operate at millikelvin temperatures in dilution refrigerators \cite{nguyen:2019prl,nguyen:2019,bhaskar:2020,stas:2022}. To avoid challenges associated with these requirements, the tin-vacancy center (SnV$^-$) in diamond has emerged as a favorable alternative due to its larger minimum ground state splitting of approximately $820 \units{GHz}$ \cite{tchernij:2017,iwasaki:2017,rugar:2019,trusheim:2020,gorlitz:2020,karapatzakis:2024}. This allows for coherent spin control at several degrees Kelvin \cite{debroux:2021,rosenthal:2023,guo:2023}, where exponentially more cooling power is available.

Recent advances of the SnV$^-$ platform include incorporation with nanophotonics \cite{rugar:2020,rugar:2021,kuruma:2021,martinez:2022,parker:2024,pasini:2023,herrmann:2023}, high-fidelity generation of single photons \cite{martinez:2022,brevoord:2024}, nuclear spin control and single-shot nuclear spin readout \cite{parker:2024} enabled by a large hyperfine coupling \cite{harris:2023}, and high-fidelity microwave spin control using both moderately strained \cite{rosenthal:2023} and highly strained \cite{guo:2023,karapatzakis:2024} centers.

However, recent work on SnV$^-$ spin control has illuminated a complicated relationship between the performance of spin polarization, coherent microwave spin control, and spin readout as functions of strain and the orientation of the applied magnetic field \cite{rosenthal:2023,guo:2023,orphalkobin:2024,pieplow:2024,karapatzakis:2024}. In particular, there is a general trade-off between optimizing for high-fidelity microwave spin control using high strain and certain magnetic field orientations, and optimizing readout using low strain and a magnetic field aligned with the spin dipole axis. To date, the only published demonstration of single-shot readout of an SnV$^-$'s electronic spin uses an unstrained center not favorable for coherent spin control \cite{gorlitz:2022}, and the only demonstrations of coherent spin control use readout far from the single-shot regime \cite{debroux:2021,rosenthal:2023,guo:2023,karapatzakis:2024}. This brings into question if rapid spin control \textit{and} high-fidelity readout are compatible for the SnV$^-$, as has previously been shown for the SiV$^-$ \cite{sukachev:2017}. 

\begin{figure}[htb!] 
\begin{center}
\includegraphics[width=1.0\columnwidth]{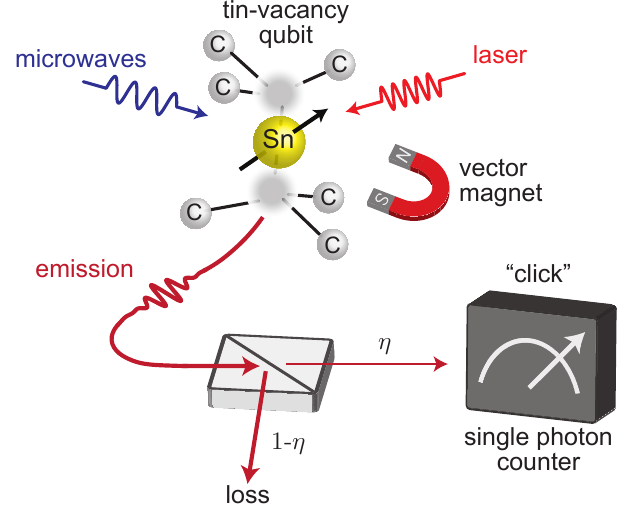}
\caption {
Schematic. A tin-vacancy (SnV$^-$) qubit consists of a tin atom replacing two carbon atoms in a diamond lattice. Here, we manipulate the spin of this atomlike system under application of a static magnetic field and using a combination of optical and microwave control pulses. This leads to spin-dependent photoluminescence, which we measure using a single-photon counter. The presence or absence of ``clicks'' on the detector is used to determine the qubit's spin state with high fidelity. Even so, some emission is lost via different channels before detection; this loss is parameterized by a beam splitter with transmittivity $\eta$ placed between the qubit and an idealized detector. Here, $\eta$ is the overall measurement efficiency.
}
\label{fig:experiment}
\end{center}
\end{figure}

In this article, by precise study of readout performance and associated trade-offs (Fig.~\ref{fig:experiment}), we answer this question favorably and further our understanding about both the SnV$^-$ platform and, in general, about the quantum measurement of solid-state emitters. We demonstrate measurement of a single SnV$^-$ electronic spin with a single-shot readout fidelity of $87.4\%$ at the same operating conditions as rapid microwave spin control. We also report a conditional readout fidelity of $98.5\%$ by conditioning on the outcome of two consecutive measurements. We achieve this performance using an overall measurement efficiency of approximately $0.1\%$, implying that near-unit fidelity is achievable in future nanophotonic devices that have much greater efficiency. Finally, we use a combination of coherent spin control and weak quantum measurement to study measurement-induced dephasing --- both affirming the basic science of quantum measurement and using the qubit's spin superposition as a resource to benchmark its interaction with light and characterize the measurement apparatus. The understanding and methods developed in this work advance SnV$^-$ based quantum technologies, enable the use of an SnV$^-$'s long spin coherence time as a resource for quantum memory, and have broad application to the study and measurement of a wide class of solid-state quantum systems.

\begin{figure*}[htb!] 
\begin{center}
\includegraphics[width=2.0\columnwidth]{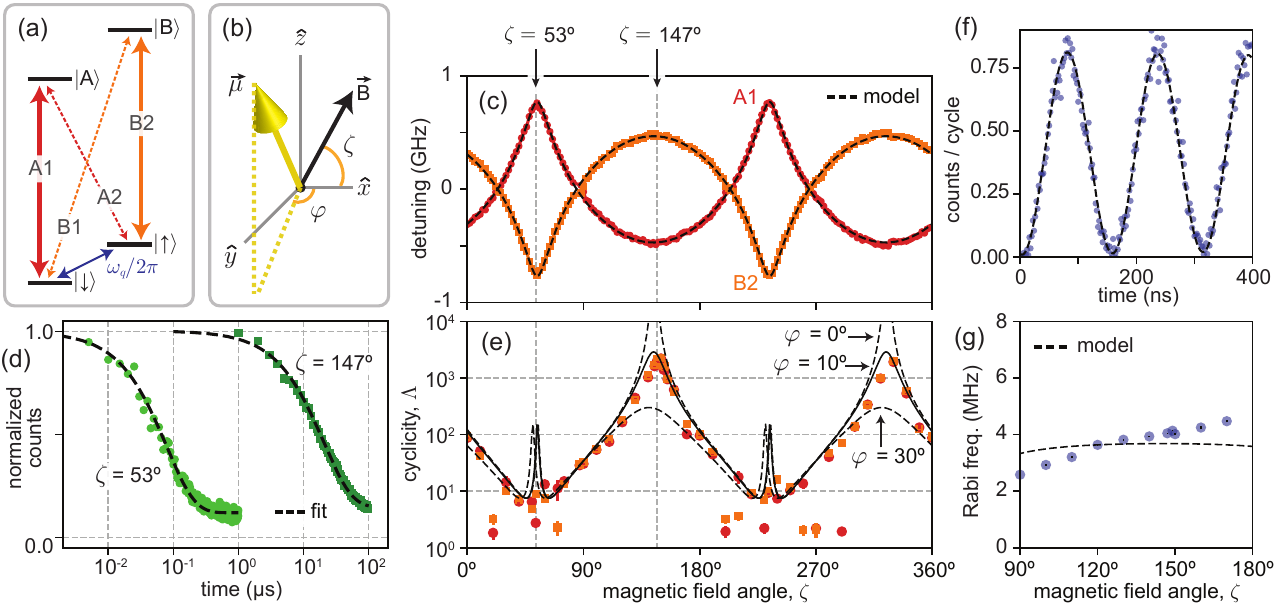}
\caption {
Spin-selective state initialization and optical and microwave control. (a) Simplified energy level diagram of the tin-vacancy qubit (SnV$^-$) in diamond. Electron spin ground states $\ket{\downarrow}$ and $\ket{\uparrow}$ are separated by the qubit frequency $\omega_q/2\pi$.  Red and orange lines illustrate optical transitions to the excited states $\ket{A}$ and $\ket{B}$, respectively. These transitions are either spin preserving (A1 and B2), or spin flipping (A2 and B1). (b) The spin, with magnetic dipole $\Vec{\mu}$, interacts with a magnetic field $\Vec{B}$ via the Zeeman effect. Here, $\Vec{B}$ is oriented at the angle $\zeta$ in the plane defined by the $\hat{x}$ and $\hat{z}$ coordinates. The spin is oriented at azimuthal angle $\varphi$ from this plane. (c) Photoluminescence excitation (PLE) measurement of the A1 and B2 transitions vs $\zeta$, fixing $|\Vec{B}|=180 \units{mT}$. Detuning is measured from 619.13972 nm (484.20808 THz). (d) Driving a spin-preserving transition polarizes the spin at rate $\Gamma_p$ (data shown for driving A1). This rate is also dependent on $\zeta$. (e) Cyclicity (the ratio of the probabilities of spin-preserving to spin-flipping decay) vs $\zeta$, determined from a fit of $\Gamma_p$ applied to Eq.~\ref{eqn:polarization_rate} in the high-power limit. Here, cyclicity changes by 3 orders of magnitude depending on $\zeta$, up to $\Lambda \approx 2.5\times10^3$. Cyclicity is expected to diverge when $\Vec{B}$ and $\Vec{\mu}$ align; here it is limited by the minimum misalignment $\varphi$, set to  $\varphi=10\degree$ for a model which follows our data (solid black line). (c) and (e) share an \textit{x} axis. (f) Rabi oscillations between the ground spin states under microwave driving, demonstrating an 80 ns $\pi$-pulse time at $\omega_q/2\pi = 3.677 \units{GHz}$. (g) Microwave Rabi frequency vs $\zeta$. At each angle, the amplitude $|\Vec{B}|$ is tuned so that $\omega_q/2\pi = 4.00 \units{GHz}$. Data are compared to a model \cite{rosenthal:2023} that uses a microwave field of amplitude 0.6 mT oriented perpendicular to $\Vec{\mu}$.
}
\label{fig:intro}
\end{center}
\end{figure*}

\section{Spin state preparation and control}
\subsection{Optical transitions}
Of crucial importance to the measurement of any qubit is its Hamiltonian. For group-IV color centers in diamond, the Hamiltonian \cite{thiering:2018} is set by both strain of the diamond lattice and by $\Vec{B}$, the external magnetic field. Here, we study the same SnV$^-$ used in Ref.~\cite{rosenthal:2023}, which has a ground state splitting of 903 GHz, larger than that of an unstrained SnV$^-$. This is in a regime of ``moderate'' strain, where the ground state strain (approximately $355 \units{GHz}$) is non-negligible but less than the spin-orbit coupling.

To understand the importance of magnetic field orientation, we first consider the SnV$^-$'s two lowest energy levels $\ket{\downarrow}$ and $\ket{\uparrow}$ in the ground state manifold and the two lowest energy levels $\ket{A}$ and $\ket{B}$ in the excited state manifold, Fig.~\ref{fig:intro}a. Both sets of levels split with magnetic field due to the Zeeman effect, which can lead to four distinct transitions as shown in Fig.~\ref{fig:intro}a: A1 and B2 (``spin-preserving''), and A2 and B1 (``spin-flipping'') \cite{hepp:2014}. The detuning between these transitions is highly dependent on the alignment between the magnetic field $\Vec{B}$ and the magnetic spin dipole $\Vec{\mu}$, parameterized by the angle $\zeta$, which is the angle of the vector $\Vec{B}$ in the plane defined by our two-axis magnet with coils along the $\hat{x}$ and $\hat{z}$ directions (see Fig.~\ref{fig:intro}b). The dipole axis is defined by the crystallographic symmetry axis that connects two missing carbon atoms and an interstitial tin vacancy (see Fig.~1). The angle $\varphi$ is a measure of misalignment between the dipole and magnetic field plane, and we have tried to minimize this angle by physical rotation of the sample.  

The A1 and B2 transitions are characterized using photoluminescence excitation (PLE) spectroscopy, Fig.~\ref{fig:intro}c, by sweeping $\zeta$ for a magnetic field of fixed amplitude $|\Vec{B}|=180\units{mT}$. Data in Fig.~\ref{fig:intro} are modeled by the Hamiltonian described in Table.~I in Ref.~\cite{karapatzakis:2024}. See Appendix~\ref{sec:fit_optical_transitions} for details, including discussion of free parameters and modifications to the SnV$^-$ Hamiltonian reported by our previous work, Ref.~\cite{rosenthal:2023}.

\subsection{Cyclicity}
In addition to setting the detuning between the A1 and B2 transitions, the magnetic field orientation controls the \textit{cyclicity}; a crucial parameter for understanding spin measurement. When pumped to $\ket{A}$ or $\ket{B}$, the SnV$^-$ subsequently relaxes to its ground state in a manner which can preserve the spin, i.e. by decay via the A1 or B2 transitions with probabilities $P_\mathrm{A1}$ or $P_\mathrm{B2}$, respectively. Alternatively, its spin can flip: i.e. by decay via the A2 or B1 transitions with probabilities $P_\mathrm{A2}$ and $P_\mathrm{B1}$, respectively. Cyclicity $\Lambda = P_\mathrm{A1}/P_\mathrm{A2} = P_\mathrm{B2}/P_\mathrm{B1}$ is the ratio of the probability of spin-preserving decay to spin-flipping decay.

Cyclicity determines how many photons are emitted, corresponding to the maximum allowed number that can be collected during readout and the optimal readout duration. For a given excitation power, a higher $\Lambda$ indicates more photons can be collected before the spin polarizes and goes ``dark'' (no longer emits photons). However, in trade-off, a longer duration of optical drive is required to polarize the spin state. A smaller $\Lambda$ will lead to faster polarization, but will limit the number of emitted photons and the useful duration of acquisition.

For example, if the SnV$^-$ is prepared in the $\ket{\downarrow}$ state given by density matrix $\hat{\rho}=\ket{\downarrow}\bra{\downarrow}$ and driven on A1 starting at time $t=0$, it will polarize into $\ket{\uparrow}$ at rate $\Gamma_p$ such that $\hat{\rho}(t)=e^{-\Gamma_p t}\ket{\downarrow}\bra{\downarrow} + (1-e^{-\Gamma_p t})\ket{\uparrow}\bra{\uparrow}$. Here,
\begin{equation}
    \Gamma_p = \frac{R}{\Lambda+1}     \label{eqn:polarization_rate}     
\end{equation}
is the polarization rate and
\begin{equation}
    R = \frac{\gamma}{2} \left(\frac{p/p_\mathrm{sat}}{1+p/p_\mathrm{sat}+(\frac{\delta}{\gamma / 2})^2}\right)
    \label{eqn:R}
\end{equation}
captures an effective optical pumping rate to $\ket{A}$, which depends on drive power $p$, on-resonance saturation power $p_\mathrm{sat}$, and detuning $\delta$ of the drive from the transition. Equation~\ref{eqn:polarization_rate} follows from the optical Bloch equations \cite{foot:2005}; see Appendix~\ref{sec:master_equation} for details. For the SnV$^-$, $\gamma\approx(4.5\pm0.2 \units{ns})^{-1}=2\pi \times (35\pm1.6\units{MHz})$ is the optical decay rate \cite{trusheim:2020}.

Intuitively, at $p/p_\mathrm{sat} \gg 1$ and $\delta \lesssim \gamma$, the rate of excitation to the higher state is much faster than relaxation; thus, $\Gamma_p \approx \gamma/2\Lambda$ is independent of power but is proportional to the saturated optical decay rate and inversely proportional to cyclicity. At $p/p_\mathrm{sat} \ll 1$, however, the excitation rate is much slower than relaxation, and $\Gamma_p \approx (\gamma / 2 \Lambda) \times (p/ p_\mathrm{sat})$ is also proportional to power.

We experimentally measure $\Gamma_p$ by binning the time-tagged photon counts collected during resonant excitation. The number of collected counts decays exponentially: $(a-b) e^{-\Gamma_p t} + b$, and data in Fig.~\ref{fig:intro}d are fit to this exponential model to determine $\Gamma_p$. Here, $t=0$ is the onset of excitation, $a$ is the maximum count rate, and $b$ is the mean count rate at $t\gg0$. The term $(a-b)e^{-\Gamma_p t}$ captures signal from SnV$^-$ emission, and the background $b$ arises from noise dominated by unwanted scatter of excitation light into collection and detector dark counts. Cyclicity $\Lambda \approx \gamma/2\Gamma_p$ is determined from $\Gamma_p$ via Eq.~\ref{eqn:polarization_rate}. For data in Fig.~\ref{fig:intro} we operate at negligible detuning (the drive frequency is recalibrated at each point) and at powers far above saturation.

Both $\Gamma_p$ and $\Lambda$ are highly dependent on the alignment between $\Vec{B}$ and $\Vec{\mu}$. This is characterized in Fig.~\ref{fig:intro}e by measuring $\Lambda$ vs the angular orientation of $\Vec{B}$ along angle $\zeta$. Cyclicity changes by 3 orders of magnitude during this sweep. At $\zeta=147\degree$, near where alignment is maximum, we measure $\Gamma_p = (20.4 \pm 0.30 \units{\mu s})^{-1} = 49.02 \pm 0.72 \units{kHz}$ corresponding to $\Lambda=2244\pm108$. At similar power and at $\zeta=53\degree$, where alignment is near minimum, we measure $\Gamma_p = (78.5\pm0.002\units{ns})^{-1} = 12.74 \pm 0.0004 \units{MHz}$ and determine $\Lambda = 8.6 \pm 0.4$. The difference between $\zeta$ values corresponding to maximum and minimum cyclicity is $94\degree$, differing from the expected value of exactly $90\degree$; we postulate that this is due to slight miscalibration of one or both magnetic coils, such that amplitude $|\Vec{B}|$ is also changing slightly with $\zeta$.

When $\Vec{B}$ and $\Vec{\mu}$ are perfectly aligned, spin-flipping relaxation becomes entirely forbidden, and cyclicity is expected to diverge ($\varphi=0$ model in Fig.~\ref{fig:intro}e). The extent to which it does not indicates remaining misalignment; e.g., the spin dipole $\Vec{\mu}$ is rotated by azimuthal angle $\varphi$ relative to the plane defined by the axes $\hat{x}$ and $\hat{z}$, Fig.~\ref{fig:intro}b. Our data follow a model using a misalignment of $\varphi=10\degree$ (solid black line, Fig.~\ref{fig:intro}e). This suggests that cyclicity can be increased by fine-tuning sample rotation and/or by using a three-axis vector magnet.

Cyclicity is minimal when $\Vec{B} \cdot \Vec{\mu} \approx 0$. However, the models in Fig.~\ref{fig:intro}e (based on the model in Ref.~\cite{rosenthal:2023}) show a slight increase in cyclicity around maximal misalignment. This feature is due to the redefinition of the spin's quantization axis along the direction of strain. Finally, we notice that for some of the angles the data differ from the model where points of lowest cyclicity occur near the angles where the A1 and B2 transitions become degenerate. We postulate that this effect could be due to a coupling interaction between these transitions indicative of a previously unexplored term in the SnV$^-$ Hamiltonian; further investigation is necessary to understand this phenomenon.

\subsection{Spin polarization fidelity}

Measurement of spin polarization, as in Fig.~\ref{fig:intro}d, also bounds the polarization fidelity. This quantifies the extent to which a resonant drive prepares the qubit in the pure states $\ket{\downarrow}$ or $\ket{\uparrow}$, and is defined as $F_\mathrm{pol} = 1 - b/(2a)$, where $a$ is the maximum count rate and $b$ is the mean count rate in the $t \gg 0$ limit \cite{trusheim:2020,debroux:2021}. We measure $F_\mathrm{pol} = 92.8\%$ at $\zeta=147\degree$ and $F_\mathrm{pol} = 94.9\%$ at $\zeta=53\degree$, consistent with other examples of high-fidelity spin polarization of SnV$^-$ centers \cite{trusheim:2020,debroux:2021,gorlitz:2022,rosenthal:2023,guo:2023}. This is limited by nonzero $b$, which is dominated by scattering from the excitation laser into the collection path. 

This technical source of noise can be mitigated by better spectral filtering, and, thus, our polarization fidelity measurement is only a lower bound. With better filtering, we would expect to bound polarization fidelity at $F_\mathrm{pol} \gtrsim 99\%$, consistent with other SnV$^-$ results \cite{trusheim:2020,debroux:2021,gorlitz:2022}. Polarization fidelity is eventually limited by the spectral overlap of the A1 and B2 transitions, which here are detuned by 500 MHz, an order of magnitude greater than their linewidths. It is also limited by the finite spin $T_1$ time, which at 1.7 K is order seconds \cite{rosenthal:2023,guo:2023}, much greater than the order 10 $\units{\mu s}$ polarization time.

\subsection{Spin control}

We now characterize performance of the SnV$^-$ as a spin qubit. When operated at maximum cyclicity ($\zeta=147\degree$) we demonstrate high-fidelity spin manipulation using a microwave control pulse, Fig.~\ref{fig:intro}f. At this operating point we demonstrate a $\pi$-pulse time of 80 ns, operated at $\omega_q/2\pi=3.677\units{GHz}$ (with $|\Vec{B}|=125 \units{mT}$) and using microwave input power and packaging similar to Ref.~\cite{rosenthal:2023}. 

In Fig.~\ref{fig:intro}g we also measure the Rabi frequency (defined as the inverse of twice the $\pi$-pulse time), as a function of $\zeta$. To avoid frequency-dependent changes to microwave power delivery, we keep the qubit frequency fixed at $\omega_q/2\pi = 4.000 \pm 0.002 \units{GHz}$ by changing the amplitude $|\Vec{B}|$ at each value of $\zeta$. Data are compared to a model (from Ref.~\cite{rosenthal:2023}) that assumes a drive of $|\Vec{B}| = 0.6 \units{mT}$ at the spin location, and which is oriented perpendicular to $\Vec{\mu}$. For this drive orientation, the Rabi frequency is expected to be highest near approximately $147\degree$ where $\Vec{B}$ and $\Vec{\mu}$ are maximally aligned, and symmetric with angle around this operating point. Disagreement between theory and experiment may be due to drifting microwave power delivery and/or miscalibration of one or both magnetic coils. However, the main result of Fig.~\ref{fig:intro}g is that microwave Rabi frequency changes little with $\zeta$. This result is consistent with numerical simulations of the SnV$^-$ Hamiltonian (Fig.~\ref{fig:tradeoff}) and shows that for this system, high cylicity and optimal readout are not to the detriment of spin control.

In summary, we have determined that resonant excitation polarizes the SnV$^-$ spin state while inducing photon emission. These effects have a strong dependence on the alignment between the spin dipole and external magnetic field, with closer alignment leading to slower polarization, higher cyclicity, and greater emission. We also demonstrate that high cyclicity is compatible with rapid microwave spin control. For the remainder of this work, we operate at $\zeta=147\degree$, where cyclicity is maximized for this experimental setup.

\section{Readout characterization}

\begin{figure*}[htb!] 
\begin{center}
\includegraphics[width=2.0\columnwidth]{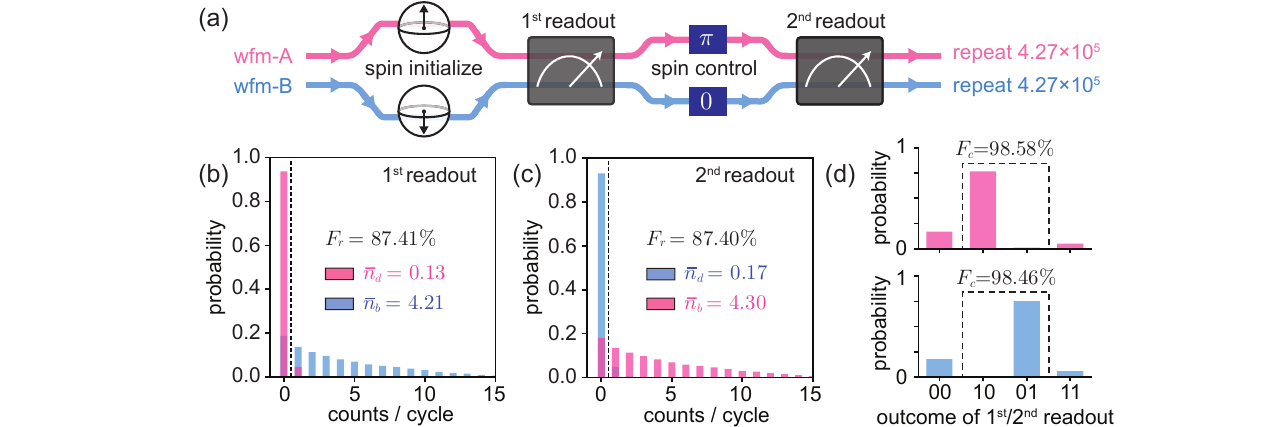}
\caption {
Single-shot readout of a single electron spin. (a) Readout is characterized using two waveforms (wfm-A and wfm-B) which are each repeated for $4.27\times10^5$ cycles. Each cycle begins with preparation of the spin in the polarized states $\ket{\uparrow}$ or $\ket{\downarrow}$ (wfm-A or wfm-B, respectively). The spin state is then projectively measured twice. Between measurements, the spin is either coherently flipped by a microwave $\pi$-pulse (wfm-A), or, no pulse is applied (wfm-B, where readout of the spin-down state already flips the spin). Each cycle also includes two charge resonance checks (CRCs; not shown --- see Fig.~\ref{fig:readout_timing} for details). (b),(c) Count distributions of the first and second readouts, respectively. Lower (higher) number distributions have a mean count number of $\bar{n}_d$ ($\bar{n}_b$) and are associated with the projection of the qubit in the $\ket{\uparrow}$ ($\ket{\downarrow}$) states prior to measurement. A given readout event yields the result ``1'' (``0'') if the counts collected during this event fall below (at or above) a threshold $N_r=1$, illustrated here by the black dashed lines. Here, both readout steps have a fidelity of $F_r=87.4\%$ (Eq.~\ref{eqn:Fr}) and (d) a conditional fidelity of $F_c\approx98.5\%$ (Eq.~\ref{eqn:Fc}). Note that data in (b)-(d) are post-selected for cycles which pass both CRCs (7.7\% of cycles for wfm-A, and 7.4\% of cycles for wfm-B, see Fig.~\ref{fig:crc} for details).
}
\label{fig:readout}
\end{center}
\end{figure*}

\subsection{Readout fidelity}
Given our understanding of spin polarization and cyclicity, we now characterize readout of the SnV$^-$ spin state. To do so, we measure the time-tagged photon statistics of two subsequent readout steps, Fig.~\ref{fig:readout}a. Depending on spin state preparation, the histogram of counts collected over many readout events in a set window follows either a ``bright'' distribution with a mean photon number of $\bar{n}_b > 0$ or a ``dark'' distribution with a mean photon number $\bar{n}_d \approx 0$. 

Readout fidelity is defined as \cite{robledo:2011}
\begin{equation}
    F_r = 1 - \frac{1}{2}P(1|\downarrow) - \frac{1}{2}P(0|\uparrow).
    \label{eqn:Fr}
\end{equation}
Here, $P(0|\uparrow)$ is the error probability of the measurement outcome ``0'' (down state), when the qubit had been initialized in $\ket{\uparrow}$. Similarly, $P(1|\downarrow)$ is the error probability of the measurement outcome ``1'' (up state), when the qubit had been initialized in $\ket{\downarrow}$. These errors are determined by the number of counts during a given readout event falling to either side of a discrimination threshold $N_r$, chosen to maximize $F_r$.

Readout results are shown in Figs.~\ref{fig:readout}b~and~\ref{fig:readout}c. After spin state initialization, two subsequent readout steps separated by a microwave $\pi$ pulse (``wfm-A''), or with no pulse (``wfm-B''). In each readout step, the resonant laser is pulsed for $100 \units{\mu s}$, long enough to fully polarize the spin state. However emitted counts are recorded only for the first $50 \units{\mu s}$ of this drive, since after this time the noise (predominantly scatter from the excitation laser) can exceed the emitted signal and, thus, reduce fidelity. For both the first and second readout steps we measure a readout fidelity of $F_r = 87.4\%$. This is in the ``single-shot'' regime, meaning a single projective measurement, or experimental ``shot,'' correctly determines the qubit state with this fidelity. For this measurement, dark distribution have $\bar{n}_d \leq 0.2$ counts per cycle, and bright distributions have $\bar{n}_b \approx 4$ counts per cycle.

Both before and after these readout steps, we include charge resonance checks \cite{robledo:2011} (CRCs, not shown in Fig.~\ref{fig:readout}a). Each CRC is a spin-independent check that consists of simultaneously driving both spin-preserving transitions (A1 and B2), and recording the check as a pass only if greater than a certain number of counts are collected during the check (see Appendix~\ref{sec:CRC} for details). Importantly, CRCs allow us to selectively characterize our qubit only when it is in the correct charge state and not suffering from spectral diffusion, and, thus, Eq.~\ref{eqn:R} may be applied in the limit where drive detuning is small compared to a linewidth. Cycles are also not recorded when the qubit ``blinks off''--- an erasure error that can be caused by the excitation laser changing the defect's charge state. By setting a CRC pass threshold to maximize readout fidelity while keeping a significant fraction of cycles, we retain $7.4\%$ ($7.7\%$) of cycles of wfm-A (wfm-B), out of $4.27\times10^5$ cycles total for each waveform. The data shown in Fig.~\ref{fig:readout} include passing cycles, only. See Fig.~\ref{fig:readout_vs_N} for analysis of readout as a function of pass threshold. 

\subsection{Conditional fidelity}
Readout can be further characterized by a ``conditional fidelity'' $F_c$, which conditions on the results of two subsequent measurements. For example, consider the characterization sequence described in Fig.~\ref{fig:readout}a. The result of these two measurements should be anticorrelated due to the insertion of the spin $\pi$ pulse during the second readout, enabling the extraction of the conditional fidelity. We, therefore, expect results ``1'' then ``0'' for wfm-A, and results ``0'' then ``1'' for wfm-B. Using the data in Fig.~\ref{fig:readout}d we find a conditional fidelity of $F_c=98.5\%$, defined as \cite{kindem:2020,lai:2024,hesselmeier:2024}
\begin{equation}
    F_c = \frac{P(j|i)}{P(i|j) + P(j|i)}.
    \label{eqn:Fc}
\end{equation}
Here, $i,j\in\{0,1\}$ are measurement results with $i \neq j$, and $P(j|i) \geq P(i|j)$ are the conditional probabilities of result $i$ ($j$) in the second readout given result $j$ ($i$) in the first readout. Like CRCs, conditional fidelity provides an avenue to retain data to include only select experimental cycles. This boosts the fidelity of selected measurements at the cost of a reduced rate of data acquisition.

\subsection{Quantum nondemolition fidelity}
Correlating the results of two subsequent measurements also quantifies the degree to which readout is quantum nondemolition (QND) \cite{lupascu:2007,raha:2020}. In quantum theory, a qubit which has been projectively measured should remain in its measured eigenstate \cite{preskill:1998}. This is a desirable property; for example, it is needed in quantum error correction algorithms that require an ancilla qubit to be repeatedly measured and real-time control logic to be implemented based on this measurement result \cite{abobeih:2022}. Repetitive measurement may also be used to enhance signal to noise of a measurement, including via use of a nuclear spin ancilla \cite{jiang:2009}.

However, a realistic quantum measurement is not always QND. For example, a measurement that correctly determines a color center's spin state, but in the process kicks the color center into a different charge state and, thus, out of the qubit subspace (erasure error), is \textit{not} QND. A measurement which heats the qubit, causing bit-flip errors independent of the measurement result, is also \textit{not} QND.

The SnV$^-$ qubit and many other atomlike systems have the property that, due to spin polarization under resonant drive, when the qubit is measured in the spin state associated with bright emission its eigenstate is also flipped. Therefore, the SnV$^-$ readout described in this work is not QND by the definition $F_\mathrm{QND}=\left[P(1|1) + P(0|0)\right]/2$ \cite{lupascu:2007}, where $P(1|1)$ and $P(0|0)$ are the conditional probabilities that subsequent measurements each yield the same result. However, we may define an alternate definition that captures the extent to which a qubit remains in the eigenstate it \textit{should} be in after measurement (i.e., is not subject to erasure error, or an unexpected bit-flip error). We define this QND-equivalent fidelity $F_q$ as
\begin{equation}
    F_q = \frac{1}{2} \left[ P(i|j) + P(j|i) \right],
    \label{eqn:F_q}
\end{equation}
where $i,j\in\{0,1\}$ and $i \neq j$. Here, $P(i|j)$ and $P(j|i)$ are the conditional probabilities that subsequent readout results are anticorrelated as they are expected to be for the experimental sequence in Fig.~\ref{fig:readout}a. Equation~\ref{eqn:F_q}, therefore, captures the essence of QND fidelity: the extent to which readout does not unexpectedly perturb the qubit. After all, a measurement result which deterministically flips the qubit state can, in principle, be corrected by a $\pi$ pulse to return the qubit to its measured state, and would, thus, be QND by the standard definition \cite{lupascu:2007}.

Using Eq.~\ref{eqn:F_q}, we compute $F_q = 76.3\%$ for the data in wfm-A and $F_q = 77.9\%$ for the data in wfm-B. Postmeasurement, the qubit therefore remains in its expected eigenstate a considerable fraction of the time. Infidelity is dominated by cycles in which both readouts yield a dark result, which may result from blinking of the qubit during readout even for cycles when both CRCs are passed or from the nonzero overlap of the bright and dark distributions. 

Finding ways to increase $F_q$ is an important research direction for future experiments and is key to implement many quantum error correction protocols. To this end, SiV$^-$ experiments that make use of cavity quantum electrodynamics (QED) to show high-fidelity QND readout \cite{nguyen:2019prl,nguyen:2019,bhaskar:2020,stas:2022,knaut:2024} are a promising blueprint for future SnV$^-$ experiments.

\begin{figure}[htb!] 
\begin{center}
\includegraphics[width=1.0\columnwidth]{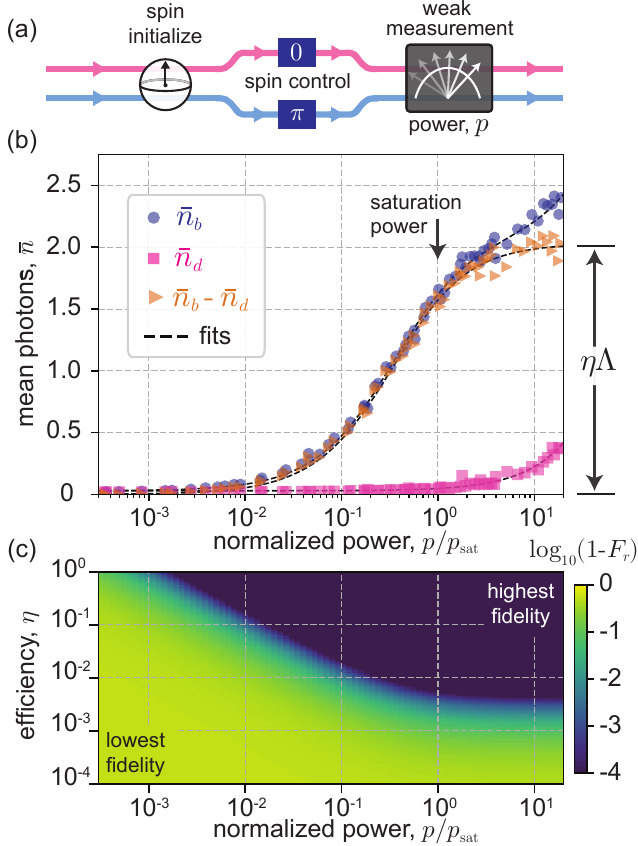}
\caption {
Measurement dependence on power and efficiency. (a) Readout performance is characterized as a function of measurement power $p$. Data are postselected on a CRC before each cycle, not shown. (b) Mean counts $\bar{n}_b$ and $\bar{n}_d$ in the bright or dark distributions, respectively, plotted vs normalized power $p/p_\mathrm{sat}$, where $p$ is the optical drive power and $p_\mathrm{sat}$ is the on-resonance saturation power. The difference $\bar{n}_b-\bar{n}_d$ captures signal from emitted photons and is fit to Eq.~\ref{eqn:emission_model} to obtain a saturation power of $p_\mathrm{sat}=313\pm8 \units{nW}$ and a measurement efficiency of $\eta=0.1\%$. (c) Simulated readout infidelity $1-F_r$, plotted on a log scale. Here, $F_r$ is simulated using Eq.~\ref{eqn:Fr_model}, which assumes emission to follow Poissonian statistics and assumes $\Lambda=2.2\times10^3$ as in experiment. Greater efficiency lowers the drive power needed to achieve high fidelity.
}
\label{fig:power}
\end{center}
\end{figure}

\subsection{Efficiency}
\label{sec:efficiency}
The electron spin readout described in Fig.~\ref{fig:readout} is not perfect, as evidenced by the overlap of the distributions in Figs.~\ref{fig:readout}b~and~\ref{fig:readout}c. In this section, we study what limits the number of collected photons, including characterizing the ``measurement efficiency'' $\eta$ that quantifies overall loss between the qubit and detector. 

Here we define measurement efficiency to be equivalent to a beam splitter with transmission $\eta$ placed between a source of emission and an ideal photon counter, Fig.~\ref{fig:experiment}. This is the probability of detecting a ``click'' per emission event and is the product of the probability of all sources of loss. It is crucial to characterize and understand $\eta$ in order to improve future experiments by reducing loss. Lower loss (increased $\eta$) will allow for higher fidelity single-shot readout and in future quantum networks will allow improved rates of entanglement generation.

Cyclicity describes the number of times, on average, that the emitter goes through excitation and relaxation before polarizing. From Figs.~\ref{fig:intro}d~and~\ref{fig:intro}e we expect $\Lambda \approx 2.2\times10^3$ emission events per readout, assuming the duration of readout is long compared to the polarization time. We measure far fewer counts per readout because most emission events are not recorded by our detector. From the qualitative argument that $\eta \approx \bar{n}_b/\Lambda$ and $\bar{n}_b \approx 4$, we estimate $\eta \approx 0.2\%$ for the data in Fig.~\ref{fig:readout}.

To further this argument, we compare the power dependence of emission to the model (Appendix~\ref{sec:master_equation}):
\begin{equation}
    \bar{n}_b - \bar{n}_d \approx \eta \Lambda (1 - e^{-\Gamma_p \tau}),
    \label{eqn:emission_model}
\end{equation}
which assumes preparation in the spin state associated with bright emission and at high power simplifies to $\bar{n}_b - \bar{n}_d \approx \Lambda \eta$. Here, $\Gamma_p$ is the power-dependent polarization rate, $\Lambda=2244\pm108$ is the cyclicity, $\tau = 50 \units{\mu s}$ is the duration of the readout collection window, and as before $\bar{n}_b$ ($\bar{n}_d$) are the mean number of counts detected when the qubit is prepared in the spin state associated with bright (dark) emission.

We fit Eq.~\ref{eqn:emission_model} to the measurement in Fig.~\ref{fig:power}b. From this fit, we determine efficiency of $\eta=0.992\times10^{-3} \pm 5.73\times10^{-6} \approx 0.1\%$; this is lower than we expect from the data in Fig.~\ref{fig:readout} because of the lower mean count number difference of $\bar{n}_b - \bar{n}_d \approx 2$ collected photons per cycle at high powers. We attribute this difference to changing conditions of the setup between measurements including drift in alignment and fluctuating polarization of the excitation laser.

Figure~\ref{fig:power}b shows that the mean number of emitted photons plateaus near a saturation power of $p_\mathrm{sat} = 313\pm8 \units{nW}$, specified at the input to the cryostat. This affirms the straightforward conclusion that readout should be optimized by operating near saturation. Below saturation, emission can be increased by pumping harder. Above saturation, noise may increase with power but signal will not. See Fig.~\ref{fig:fidelity_vs_power}a for a measurement of readout fidelity vs power.

Many sources of loss contribute to inefficiency including nonradiative decay, filtering of emission to collect the phonon sideband only, loss between the qubit and detector, and detector inefficiency (see Table~\ref{tab:loss_budget} for details). However, we estimate the dominant source of loss is the scattering of emission into bulk diamond rather than into the collection path. Although this is the same emitter used in Ref.~\cite{rosenthal:2023}, from fine-tuned optimization using nanopositioners we confirm that this emitter is actually in a mesa structure (see Fig.~\ref{fig:scattering}) rather than a photonic nanopillar. Numerical simulations of an SnV$^-$ center within this structure predict that only a modest approximately $5\%$ of emitted light is routed to collection (compared to approximately $3\%$ for an emitter in bulk). 

This fraction can be greatly increased by use of nanophotonics. For example, at visible wavelengths this number can be increased to approximately 90\% using photonic waveguides or photonic crystals \cite{riedel:2023,rugar:2021}. 
Likewise, inverse-designed grating couplers can extract approximately $25\%$ \cite{dory:2019} and tapered fiber coupling up to approximately $90\%$ \cite{burek:2017} of photons in a waveguide mode. Since stable, narrow linewidth SnV$^-$'s have already been successfully incorporated into nanophotonic structures \cite{rugar:2020,rugar:2021,kuruma:2021,martinez:2022,parker:2024,herrmann:2023,pasini:2023}, utilizing nanophotonics offers a clear path toward an order-of-magnitude higher efficiency, which will advantageously improve readout fidelity at low excitation powers, Fig.~\ref{fig:power}c.

Improved efficiency also lowers the cyclicity required for single-shot readout (generally $\eta \Lambda > 1$), allowing for single-shot readout over expanded magnetic field orientations and strain regimes. For instance, this will enable single-shot readout of highly strained SnV$^-$'s which, in general have lower cyclicity with more sensitivity to magnetic field alignment, but for which microwave spin control can require less drive power \cite{guo:2023,karapatzakis:2024}.

\begin{figure*}[htb!] 
\begin{center}
\includegraphics[width=2.0\columnwidth]{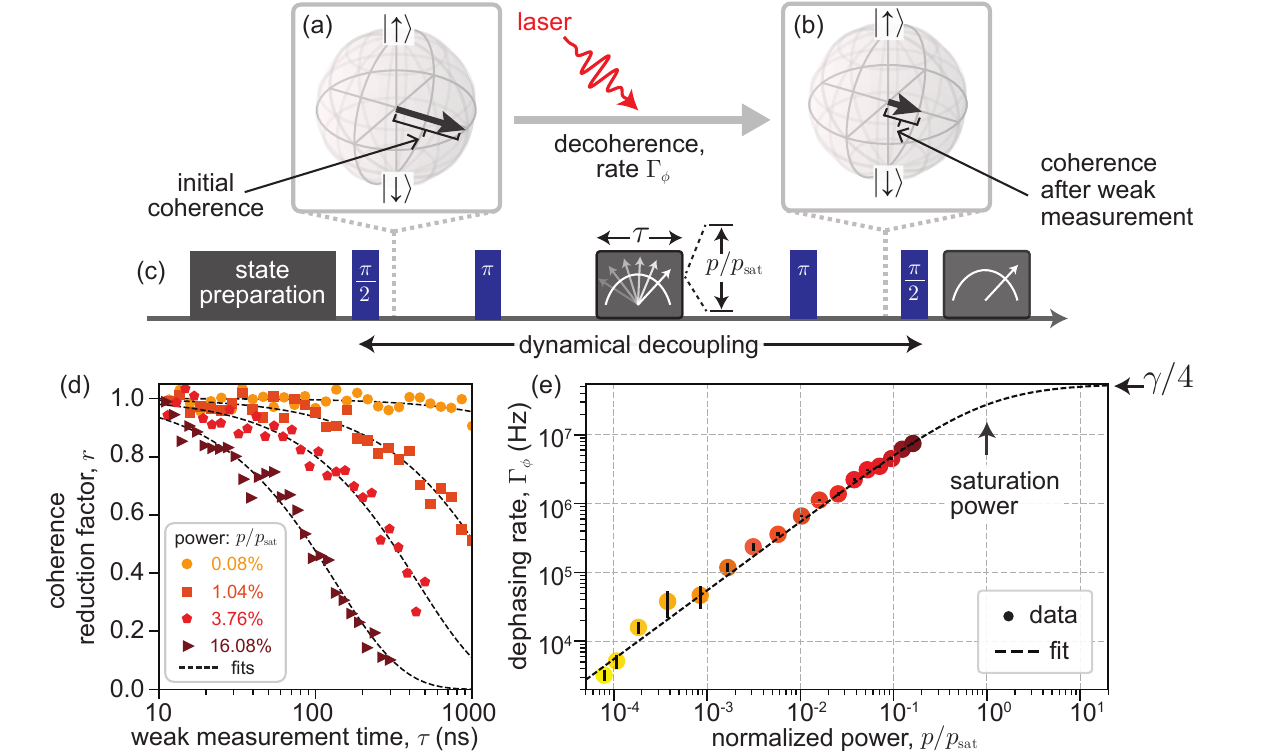}
\caption {
Control of qubit dephasing via weak quantum measurement. (a) A qubit prepared in a superposition state has an initial \textit{coherence}, specified by the amplitude of its off-diagonal density matrix elements $|\hat{\rho}_{\downarrow \uparrow}|$. This is illustrated on the Bloch sphere for the superposition state $(\ket{\downarrow} + \ket{\uparrow})/\sqrt{2}$. (b) Measurement causes decoherence at rate $\Gamma_\phi$. In a projective measurement the qubit is entirely dephased, but, more generally, after a weak measurement its coherence is reduced to $|\hat{\rho}_{\downarrow \uparrow}^\prime| \leq |\hat{\rho}_{\downarrow \uparrow}|$. (c) This interplay is studied by inserting a weak measurement of variable power $p/p_\mathrm{sat}$ and variable duration $\tau$ into the center of a dynamical decoupling experiment. (See Fig.~\ref{fig:readout_timing} for details.) The (projective) readout signal of this experiment measures the ratio $r$ of the remaining coherence to the initial coherence, $r = |\hat{\rho}_{\downarrow \uparrow}^\prime| / |\hat{\rho}_{\downarrow \uparrow}|$, which decays exponentially with weak measurement time as $r = \mathrm{exp}\left(-\Gamma_\phi \tau\right)$. (d) Selected measurements of $r$ as a function of $\tau$ and $p/p_\mathrm{sat}$ and compared to exponential fits. (e) Dephasing rate vs power. Below saturation, the dephasing rate rises linearly with power. Above saturation, it plateaus to $\gamma/4$, the maximum dephasing rate set by the SnV$^-$'s optical decay rate $\gamma$. Data are fit to Eq.~\ref{eqn:dephasing_rate}.
}
\label{fig:dephasing}
\end{center}
\end{figure*}

\section{Weak measurement of a color center qubit}
\label{sec:weak_meas}

\subsection{Controlled dephasing}
Aside from the practical use of characterizing efficiency and optimizing readout, Fig.~\ref{fig:power} illustrates that the power dependence of readout offers a broad pathway for exploring quantum measurement. Measurement is fundamental to quantum theory and is taken as one of the axioms of quantum mechanics \cite{preskill:1998}. Under the action of measurement, a quantum state undergoes probabilistic collapse into one of its eigenstates. This collapse destroys quantum superposition, i.e., causes decoherence.

After a \textit{projective} measurement, for example, as demonstrated in Fig.~\ref{fig:readout}, a measured quantum state is entirely dephased and is projected into a classical, probabilistic mixture of eigenstates. In contrast, for the case of a \textit{weak} measurement \cite{clerk:2010}, a quantum state has a finite chance of collapse and therefore, on average, retains some coherence. This equivalently limits the classical information a detector may acquire about its state. In this section, we combine the techniques we have thus far demonstrated: high-fidelity spin control (Fig.~\ref{fig:intro}), single-shot projective measurement (Fig.~\ref{fig:readout}), and weak measurement (Fig.~\ref{fig:power}) in order to study measurement-induced dephasing of the SnV$^-$'s electronic spin (Fig.~\ref{fig:dephasing}). This study is an interesting parallel to recent work in the superconducting qubit community, e.g. \cite{clerk:2010,bultink:2018,rosenthal:2021,lecocq:2021,white:2023}, and contrasts from previous color center work focused on nuclear spin control \cite{blok:2014,liu:2017,pfender:2019,cujia:2019}. This study also contrasts from cavity QED-based projective readout of color center electronic spins \cite{nguyen:2019prl,nguyen:2019,bhaskar:2020} but, importantly, has a direct analogy to the measurement-induced decoherence of a nuclear spin state in a cavity-coupled SiV$^-$, studied in Ref.~\cite{stas:2022}.

To study measurement-induced dephasing we do the experiment described in Fig.~\ref{fig:dephasing}: The qubit is prepared in a superposition state with initial coherence $|\hat{\rho}_{\downarrow \uparrow}|$, subjected to a weak measurement, and finally, projectively measured to determine its remaining postmeasurement coherence, $|\hat{\rho}_{\downarrow \uparrow}^\prime|$. The weak measurement pulse consists of a resonant laser drive of power $p$ and duration $\tau$ and is placed at the center of a dynamical decoupling sequence (see Fig.~\ref{fig:readout_timing}d for details). As $\tau$ and $p$ increase we observe a decrease in readout contrast, which is proportional to the qubit's remaining coherence after undergoing weak measurement. We fit this to the exponential model $|\hat{\rho}_{\downarrow \uparrow}^\prime| = |\hat{\rho}_{\downarrow \uparrow}| e^{-\Gamma_\phi \tau}$, where $\Gamma_\phi$ is the measurement-induced dephasing rate.

We fit the measurement of $\Gamma_\phi$ as a function of power, Fig.~\ref{fig:dephasing}d, to the model (see Appendix~\ref{sec:master_equation}):
\begin{equation}
    \Gamma_\phi = \frac{R}{2},
    \label{eqn:dephasing_rate}
\end{equation}
where $R$ is given by Eq.~\ref{eqn:R}. Notice there is a simple relationship $\Gamma_\phi \approx \Lambda \Gamma_p/2$ between Eqs.~\ref{eqn:polarization_rate}~and~\ref{eqn:dephasing_rate}. Polarization, dephasing, and emission are all closely related with power dependence governed by the relative ratio of pump power to its saturation value, $p/(p+p_\mathrm{sat})$, as follows from the optical Bloch equations, and with the overall timescale of dynamics set by the optical decay rate $\gamma$.

This experiment offers a way to precisely study how strongly the laser drive interacts with a qubit. In the low-power limit, dephasing rises linearly with $p$. Above saturation, dephasing must plateau to the limit controlled by the optical decay rate $\gamma$. A fit of Eq.~\ref{eqn:dephasing_rate} to the data in Fig.~\ref{fig:dephasing}d determines a saturation power of $p_\mathrm{sat} = 1436 \pm 21 \units{nW}$, specified at the input to the cryostat. We attribute the difference between $p_\mathrm{sat}$ here compared to the measurement in Fig.~\ref{fig:power} to drift in the optical setup between datasets and/or that the data in Fig.~\ref{fig:power} are post-selected on CRCs whereas data in Fig.~\ref{fig:dephasing} are not. 

We note that the CRC pass fraction does not change with power until $p \gg p_\mathrm{sat}$ (see Fig.~\ref{fig:fidelity_vs_power}b), and we furthermore note that an off-resonant control pulse does not cause any measurement-induced dephasing (Fig.~\ref{fig:mid_control}). These experiments indicate that the observed rise in dephasing rate with drive power is not due to an alternate effect such as drive-induced blinking or spectral diffusion, a conclusion which is consistent with prior demonstrations of coherent optical Raman spin control \cite{debroux:2021,rosenthal:2023,guo:2023}.

\subsection{A general method to characterize efficiency}
\label{sec:general_method_to_characterize_efficiency}

\subsubsection{Photoluminescent readout}

Apart from the fundamental scientific curiosity surrounding study of quantum measurement, examining the low-power limits of measurements depicted in both Fig.~\ref{fig:readout}c~and~\ref{fig:dephasing}e offers an alternative means to evaluate measurement efficiency. Far below saturation, both the mean photon count difference $\bar{n}_b - \bar{n}_d$ and the measurement-induced dephasing rate $\Gamma_\phi$ scale linearly with power. Taylor expanding Eq.~\ref{eqn:emission_model} in the $p/p_\mathrm{sat} \ll 1$ limit and for $\delta=0$ gives $\bar{n}_b - \bar{n}_d = A p + \mathcal{O}(p^2)$ with proportionality constant $A = \gamma \eta \tau / (2 p_\mathrm{sat}) = 1.744\times10^{-2} \pm 8.305\times10^{-5} \units{nW^{-1}}$. At low power, $\Gamma_\phi = B p + \mathcal{O}(p^2)$ is also linear with proportionality constant $B=\gamma/(4p_\mathrm{sat})=41.5\pm2.4 \units{kHz/nW}$. Therefore,
\begin{equation}
    \eta = \frac{A}{2 B \tau},
    \label{eqn:eta_compare_rates}
\end{equation}
where $A$ is a fit of Eq.~\ref{eqn:emission_model} to the weak measurement data in Fig.~\ref{fig:power}b, $B$ is a fit of Eq.~\ref{eqn:dephasing_rate} to the measurement-induced dephasing data in Fig.~\ref{fig:dephasing}e, and $\tau = 50 \units{\mu s}$ is the readout integration window. See Ref.~\cite{bultink:2018} for a similar approach for characterizing the efficiency of superconducting qubit measurement.

Here, solving Eq.~\ref{eqn:eta_compare_rates} gives $\eta = 0.420\% \pm 0.024\%$. This is greater than the fit to the data in Fig.~\ref{fig:power}c alone, and is consistent with the greater saturation power of the data in Fig.~\ref{fig:dephasing}d. This indicates the laser is interacting less strongly with the qubit in the dataset used to fit $B$, which we postulate is due to differences in the measurement conditions between these experiments. 

Despite this difference, we include Eq.~\ref{eqn:eta_compare_rates} because it provides a general framework that may be useful in future experiments requiring precise characterization of solid-state atomic systems. This measurement allows for determination of efficiency within the low-power limit, thus avoiding the effect of power-dependent qubit instability. It may be particularly applicable to other color centers which are difficult to saturate or have a competing ionization process.

\subsubsection{Dispersive readout}
The fundamental physics of weak quantum measurement is agnostic to technique, platform, or type of detector. The introduced framework for accurately characterizing the measurement efficiency also applies to other types of readout, including the case of dispersive measurement of cavity QED systems. Dispersive readout is an important ingredient to establishing quantum network nodes based on coherent atom-light interactions \cite{nguyen:2019,nguyen:2019prl,bhaskar:2020,stas:2022,knaut:2024}.

In dispersive readout, discrimination between the $\ket{\uparrow}$ and $\ket{\downarrow}$ states is governed by the coupling between the spin state and cavity photons, causing spin-dependent scattering of a probe tone. Weak and strong measurement regimes can also be understood systematically based on the same characterization of measurement-induced dephasing used in Fig.~\ref{fig:dephasing}, such that a quantitative comparison between the dephasing rate $\Gamma_\phi$ and the collected information about the qubit state (e.g., signal $\bar{n}_b - \bar{n}_d$) may be used to determine measurement efficiency $\eta$. See Appendix~\ref{appendix:cqed_readout} for a detailed mathematical discussion.

As with the photoluminescence-based readout discussed elsewhere in this work, efficiency is important for dispersive readout, because greater efficiency allows for high-fidelity measurement using lower probe powers and shorter readout durations, thus mitigating drive-induced sources of error. We also expect that precise characterization of efficiency will be an important tool for the development of quantum repeater nodes, in parallel to its general application to benchmarking sophisticated circuit QED systems \cite{bultink:2018,white:2023}. After all, careful measurement of loss is a first step toward its mitigation, and reducing loss will improve the entanglement generation rate within a quantum network. 

\section{Conclusion}
In conclusion, we demonstrate single-shot readout of a single electronic spin of an SnV$^-$ qubit in diamond, along with coherent control of this spin by a microwave drive. We report a readout fidelity of $F_r = 87.4\%$ using a $50 \units{\mu s}$ readout integration window. These results are achieved by aligning the magnetic field near the spin dipole direction (misalignment of $\varphi\approx10\degree$) to operate at cyclicity $\Lambda \approx 2200$. To our knowledge, this is highest fidelity single-shot readout of an SnV$^-$ spin that has been published to-date: The readout in Ref.~\cite{gorlitz:2022} reports a lower fidelity of $74\%$, and uses an unstrained emitter for which microwave spin control is not feasible. (Reference~\cite{parker:2024} reports readout of an SnV$^-$ nuclear spin and also does not include spin control.) Next, using high-fidelity readout, we repeatedly measure the qubit and characterize the extent to which subsequent measurements are correlated; doing so, we characterize a conditional fidelity of $F_c = 98.5\%$ (Eq.~\ref{eqn:Fc}), and a QND-equivalent fidelity of $F_q \approx 77\%$. All of these results are obtained operating at a measurement efficiency of only $\eta \approx 0.1\%$, indicating orders-of-magnitude room for improvement in future devices that utilize nanophotonics for higher collection efficiency.

Finally, we use rapid microwave spin control and long spin coherence to study dephasing induced by measurement. This demonstrates the fundamental interplay between measurement and dephasing that is inherent to quantum mechanics. It is also a  versatile technique to characterize how strongly a laser drive interacts with a qubit and to characterize measurement efficiency. As an immediate next step, it will be of interest to apply the readout techniques developed in this manuscript to dispersive readout of cavity QED systems. In direct analogy with its use in circuit QED \cite{bultink:2018,rosenthal:2021,lecocq:2021,white:2023,blais:2021}, studying the spin-dependent interaction of a probe tone with a cavity will be a useful way to benchmark measurement efficiency and understand sources of loss.

Overall, our work advances the SnV$^-$ as a platform for building quantum networks. Taken together with other recent SnV$^-$ work including nanophotonic integration \cite{rugar:2021,kuruma:2021,herrmann:2023,pasini:2023}, microwave spin control \cite{rosenthal:2023,guo:2023}, spin-photon entanglement \cite{martinez:2022}, and nuclear spin states \cite{harris:2023,parker:2024}, the SnV$^-$ has all the features necessary for building a quantum repeater node of similar scale and complexity to SiV$^-$-based devices \cite{nguyen:2019prl,nguyen:2019,bhaskar:2020,stas:2022} but which favorably are robust to heating effects and elevated temperatures. More broadly, the understanding and methods we develop here--- in particular, the use of weak quantum measurement and measurement-induced dephasing---serve as powerful metrological tools and are applicable to the study of a wide variety of solid-state atomlike systems.

\vspace{0.1in}
While in press, the authors became aware of relevant work in Ref.~\cite{beukers:2024}.

\vspace{0.1in}
\textit{Acknowledgments.}--- This work has been supported by the Department of Energy under the Q-NEXT program and Grant No. DE-SC0020115 and by the National Science Foundation Grant No. ECCS 2150633. E.I.R. and C.P.A. acknowledge support by an appointment to the Intelligence Community Postdoctoral Research Fellowship Program at Stanford University administered by Oak Ridge Institute for Science and Education (ORISE) through an interagency agreement between the U.S. Department of Energy and the Office of the Director of National Intelligence (ODNI). J.C. acknowledges support from the Terman Faculty Fellowship at Stanford. H.C.K. acknowledges support by the Burt and Deedee McMurtry Stanford Graduate Fellowship (SGF). J.G. acknowledges support from the Hertz Fellowship. G.S. and S.A. acknowledge support from the Stanford Bloch Postdoctoral Fellowship. D.R. acknowledges support from the Swiss National Science Foundation (Project No. P400P2\_194424). D.R. and S.A. contributed to this work prior to joining AWS. This work was performed, in part, at the Center for Integrated Nanotechnologies, an Office of Science User Facility operated for the U.S. Department of Energy (DOE) Office of Science. Sandia National Laboratories is a multimission laboratory managed and operated by National Technology \& Engineering Solutions of Sandia, LLC, a wholly owned subsidiary of Honeywell International, Inc., for the U.S. DOE’s National Nuclear Security Administration under Contract No. DE-NA-0003525. 

The views expressed in the article do not necessarily represent the views of the U.S. DOE or the United States Government. We thank Jacob Feder, Jonathan Marcks, Mia Froehling Gallier, and Cyrus Zeledon for help with instrument control code, based on the ``nspyre'' framework \cite{feder:2023}. We thank Haiyu Lu, Shuo Li, Patrick McQuade, Zhi-Xun Shen and Nicholas Melosh for assistance with diamond sample preparation. We thank Nazar Delgan and F. Joseph Heremans for collaboration on the preparation of related devices. We thank Noah Lordi for helpful discussions. Part of this work was performed at the Stanford Nanofabrication Facility (SNF) and the Stanford Nano Shared Facilities (SNSF), supported by the NSF under Grant No. ECCS-2026822. 

\appendix
\label{appendix}

\section{Readout model}
\subsection{Dynamics of emission, polarization and dephasing}
\label{sec:master_equation}
\subsubsection{Three-level model}
\label{sec:3levels}
\begin{figure}[htb!] 
\begin{center}
\includegraphics[width=1.0\columnwidth]{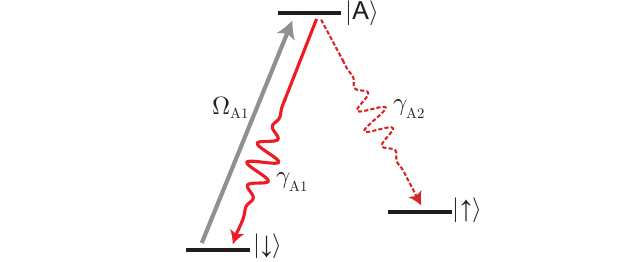}
\caption{
A three-level atomic system is driven to the $\ket{A}$ state at rate $\Omega_\mathrm{A1}$. From $\ket{A}$, the system emits radiation via relaxation to the $\ket{\downarrow}$ state at rate $\gamma_\mathrm{A1}$, or the $\ket{\uparrow}$ state at rate $\gamma_\mathrm{A2}$.
}
\label{fig:lambda}
\end{center}
\end{figure}

In this section, we model the dynamics of an SnV$^-$ center under a resonant drive. This is used to model spin polarization, photon emission during readout, and measurement-induced dephasing. Consider a three-level system (a ``lambda system'') consisting of two ground states $\ket{\downarrow}$ and $\ket{\uparrow}$, with an excited state $\ket{A}$, Fig.~\ref{fig:lambda}. This is a simplified version of the energy diagram in Fig.~\ref{fig:intro}a. Relaxation from $\ket{A}$ occurs with probability $P_\mathrm{A1}$ or $P_\mathrm{A2}$ into the $\ket{\downarrow}$ or $\ket{\uparrow}$ states, respectively, and cyclicity $\Lambda=P_\mathrm{A1} / P_\mathrm{A2}$. Since $P_\mathrm{A1} + P_\mathrm{A2} = 1$ we have $P_\mathrm{A2} = 1 /(1+\Lambda)$ and $P_\mathrm{A1} = \Lambda / (1+\Lambda)$.

We model this system using the master equation (setting $\hbar=1$):
\begin{align}
&\frac{d}{dt}\hat{\rho} = -i[ \hat{H},\hat{\rho}] + \mathfrak{L}[\hat{\rho}], \label{eqn:master_equation} \\
&\mathfrak{L}[\hat{\rho}] = \sum_{k=1}^2 \hat{L}_k \hat{\rho} \hat{L}_k^\dag - \frac{1}{2}(\hat{L}_k^\dag \hat{L}_k \hat{\rho} + \hat{\rho} \hat{L}_k^\dag \hat{L}_k).
\label{eqn:loss_operators}
\end{align}
Here, $\hat{\rho}$ is the density matrix of the three-level system in Fig.~\ref{fig:lambda}, $\hat{H}$ is the Hamiltonian, and $\hat{L}_k \in \{\hat{L}_1, \hat{L}_2 \}$ are the stochastic quantum jump operators.

Under a Rabi drive the Hamiltonian in a rotating frame is
\begin{align}
    &\hat{H} = \frac{\Omega_\mathrm{A1}}{2}\Bigl( \ket{\downarrow}\bra{A} + \ket{A} \bra{\downarrow} \Bigr) + 
    \frac{\delta}{2}\Bigl( \ket{\downarrow}\bra{\downarrow} - \ket{A} \bra{A} \Bigr),
    \label{eqn:hamiltonian}
\end{align}
where the first term contains the optical Rabi frequency $\Omega_\mathrm{A1}$, and the second term describes its detuning $\delta$ from the A1 transition. We also define stochastic quantum jump operators:
\begin{align}
    &\hat{L}_1 = \sqrt{\gamma_\mathrm{A1}} \ket{\downarrow} \bra{A}, 
    \label{eqn:collapse_operators1}
    \\
    &\hat{L}_2 = \sqrt{\gamma_\mathrm{A2}} \ket{\uparrow} \bra{A}, \label{eqn:collapse_operators2}
\end{align}
with $\gamma_\mathrm{A1} = \gamma \Lambda/(1+\Lambda)$ and $\gamma_\mathrm{A2} = \gamma/(1+\Lambda)$, and where $\gamma = \gamma_\mathrm{A1} + \gamma_\mathrm{A2}$ is the optical decay rate given by the lifetime of the $\ket{A}$ state.

\subsubsection{Two-level model}
\label{sec:2levels}
This model can be simplified by considering only the subspace spanned by the spin levels $\ket{\downarrow}$ and $\ket{\uparrow}$. This obfuscates the coherent dynamics between the spin ground states and the $\ket{A}$ state, but captures the effects of spin polarization and measurement-induced dephasing.

With this simplification, the master equation becomes $\frac{d}{dt}\hat{\rho} = \mathfrak{L}[\hat{\rho}]$ with operators:
\begin{align}
    &\ \hat{L}_1 = \sqrt{R_1} \ket{\downarrow} \bra{\downarrow}, \label{eqn:collapse_operators1_simplified} \\ &\ \hat{L}_2 = \sqrt{R_2} \ket{\uparrow} \bra{\downarrow},
    \label{eqn:collapse_operators2_simplified}
\end{align}
only. Here, rates $R_1=R P_\mathrm{A1}= R \Lambda/(1+\Lambda)$ and $R_2 = R P_\mathrm{A2} = R / (1+\Lambda)$ are
\begin{equation}
    R = \frac{\gamma}{2} \left(\frac{p/p_\mathrm{sat}}{1+p/p_\mathrm{sat}+(\frac{\delta}{\gamma / 2})^2}\right),
    \label{eqn:Rappendix}
\end{equation}
where $p/p_\mathrm{sat} = 2\Omega_\mathrm{A1}^2/\gamma^2$ is the resonant saturation parameter, which can be derived from the optical Bloch equations \cite{foot:2005}. 

These dynamics have a clean analytical solution. When driven on the A1 transition, a qubit starting at time $t=0$ in the state given by density matrix $\hat{\rho}=\ket{\psi}\bra{\psi}$, with $\ket{\psi}=\alpha \ket{\downarrow} + \beta \ket{\uparrow}$ and $|\alpha|^2+|\beta|^2=1$, will evolve as:
\begin{align}
    &\ \rho_{\downarrow \downarrow}[t] = |\alpha|^2 e^{-\Gamma_p t}, \label{eqn:rho11_analytic}
    \\
    &\ \rho_{\uparrow \downarrow}[t] = \alpha \beta^* e^{-\Gamma_\phi t},  \label{eqn:rho12_analytic}
    \\
    &\ \rho_{\downarrow \uparrow}[t] = \alpha^* \beta e^{-\Gamma_\phi t},  \label{eqn:rho21_analytic}
    \\    
    &\ \rho_{\uparrow \uparrow}[t] = 1 + (|\beta|^2 - 1) e^{-\Gamma_p t},  \label{eqn:rho22_analytic}
\end{align}
where $\Gamma_\phi = R/2$ and $\Gamma_p = R/(1+\Lambda)$. The qubit, therefore, undergoes drive-induced dephasing at rate $\Gamma_\phi$, and polarization at the slower rate $\Gamma_p$.

\subsubsection{Comparison between models}
To confirm these models, we compare simulations of the three-level model in Appendix~\ref{sec:3levels} to the two-level model in Appendix~\ref{sec:2levels}. In each case, we numerically solve the master equation using the QuTiP package \cite{johansson:2012}.

\begin{figure}[htb!] 
\begin{center}
\includegraphics[width=1.0\columnwidth]{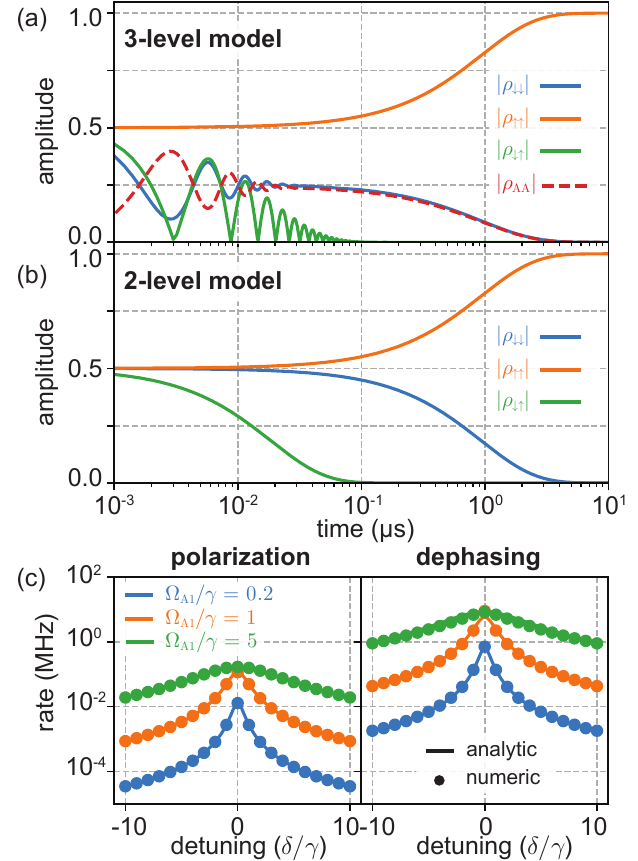}
\caption{
(a) Dynamics of a driven three-level atomic system, Fig.~\ref{fig:lambda}, simulated using the Hamiltonian and collapse operators given by Eqs.~\ref{eqn:hamiltonian}-\ref{eqn:collapse_operators2}. (b) Dynamics of the simplified two-level model, simulated using collapse operators Eqs.~\ref{eqn:collapse_operators1_simplified}~and~\ref{eqn:collapse_operators2_simplified}. Both (a) and (b) use $\Lambda=100$, $\gamma/2\pi = 35 \units{MHz}$, $\Omega_\mathrm{A1} = 5\gamma$, and $\delta=0$. (c) Polarization rate $\Gamma_p$ (left) and dephasing rate $\Gamma_\phi$ (right) as functions of $\delta$ and at different values of $\Omega_\mathrm{A1}/\gamma$. Numerical solutions to the three-level model described in Appendix~\ref{sec:3levels} (dots) match the analytical solutions (Eqs.~\ref{eqn:rho11_analytic}-\ref{eqn:rho22_analytic}) based on the two-level model in Appendix~\ref{sec:2levels} (lines).
}
\label{fig:master_equation}
\end{center}
\end{figure}

The time-domain dynamics of both models are plotted in Fig.~\ref{fig:master_equation}. In both cases, the system undergoes spin polarization at rate $\Gamma_p$ and an overall envelope of drive-induced dephasing at rate $\Gamma_\phi$. The three-level model also captures the coherent dynamics between the spin ground states and the $\ket{A}$ state. We simulate both models as a function of detuning $\delta$ and confirm they match. This comparison also illustrates power broadening---the effective increase in linewidth at high drive powers.

\subsection{Emission model}
\label{sec:model_for_emission}
A driven atomic system emits at the steady-state photon scattering rate given by Eq.~\ref{eqn:Rappendix}, times the population in the driven spin state \cite{foot:2005}. For example, upon onset of a laser drive at time $t=0$, the emission rate is
\begin{align}
    \Gamma_\mathrm{em}(t) = |\alpha|^2 R e^{-\Gamma_p t},
    \label{eqn:Gamma_em}
\end{align}
where $|\alpha|^2$ is the initial population in the $\ket{\downarrow}$ state. We now use this rate to model readout. 

In readout, ``dark'' or ``bright'' distributions with low ($\bar{n}_d$) or high ($\bar{n}_b$) mean photon numbers are associated with the qubit being prepared in different spin states prior to measurement. For example, state preparation of $\ket{\uparrow}$ yields $\alpha=0$, no initial emission, and a dark collected distribution. State preparation of $\ket{\downarrow}$ yields $\alpha=1$, maximum initial emission, and a bright collected distribution. However, emitted photons are collected only with a probability $\eta$ (the measurement efficiency), and, furthermore, collected counts can also come from unwanted sources unrelated to the qubit, i.e., noise arriving at rate $\Gamma_\mathrm{noise}$. We, therefore, model $\bar{n}_d$ and $\bar{n}_b$ as
\begin{align}
    &\bar{n}_d = \int_0^\tau \Gamma_\mathrm{noise} dt, \label{eqn:nd}
    \\
    &\bar{n}_b = \int_0^\tau \left(\Gamma_\mathrm{noise} + \eta \Gamma_\mathrm{em} \right) dt,
    \label{eqn:nb}
\end{align}
respectively, where $\tau$ is the readout integration window. We model noise to be time independent arriving at rate $\Gamma_\mathrm{noise} = a + b p$, where $a$ is the mean rate of background noise photons including dark counts, $b$ is noise from scatter of the excitation laser into the collection path and is proportional to power.

For preparation in the spin state associated with maximum emission ($\alpha=1$), the difference in collected counts between the bright and dark distributions is, therefore, $\bar{n}_b - \bar{n}_d = \eta (\Lambda + 1) (1 - e^{-\Gamma_p \tau})$. In Eq.~\ref{eqn:emission_model}, we simplify this by taking $(\Lambda+1) \to \Lambda$, since in this work we operate in the $\Lambda\gg1$ limit where $(\Lambda+1)\approx\Lambda$. For $\Lambda \tau \gg 1$, so the duration of drive is long enough to entirely polarize the spin state, this reduces to $\bar{n}_b - \bar{n}_d \approx \eta \Lambda$. In other words, the expected number of detection events is essentially the cyclicity $\Lambda$ times the measurement efficiency $\eta$.

\subsection{Readout fidelity vs power}
\label{sec:Fr_vs_power}
We now use this model of emission to study readout fidelity as a function of power. Readout fidelity quantifies the distinction between the bright or dark distributions of photons collected when the qubit is prepared in the $\ket{\downarrow}$ or $\ket{\uparrow}$ spin state. These distributions are distinguished by a readout threshold $N_r$, meaning the outcome ``0'' can be associated with the collection of $\geq N_r$ counts per readout window and the outcome ``1'' with $< N_r$ counts. In practice, $N_r$ is chosen to maximize readout fidelity $F_r=1-\frac{1}{2}P(1|\downarrow)-\frac{1}{2}P(0|\uparrow)=\frac{1}{2}P(0|\downarrow)+\frac{1}{2}P(1|\uparrow)$. 

Here, $P(1|\downarrow)$ and $P(0|\uparrow)$ are error probabilities, associated with the chance a collected number of counts falls within a different distribution than expected given state preparation. Infidelity can result from too many counts in the dark distribution or too few counts in the bright distribution. It can also result from state preparation errors $\epsilon_1$ ($\epsilon_0$), the chances the qubit was actually prepared in the $\ket{\downarrow}$ ($\ket{\uparrow}$) state when it was attempted to be prepared in the $\ket{\uparrow}$ ($\ket{\downarrow}$) state, respectively.

To model readout fidelity, we assume that emission follows Poissonian statistics arising from a Markovian process. These statistics are known to describe emission from atomic systems \cite{danjou:2014} but can be modified in other experiments where interaction with a photonic structure changes the electromagnetic density of states \cite{pasini:2023}. Poissonian statistics are captured by the model in Eq.~\ref{sec:master_equation}, which leads to the emission rate in Eq.~\ref{eqn:Gamma_em}. Therefore, during a readout event, collection of $k$ counts is expected with probability
\begin{equation}
    p[k,n] = \frac{n^k e^{-n}}{k!},
    \label{eqn:poisson}
\end{equation}
where $n=\bar{n}_b$ ($n=\bar{n}_d$) for readout associated with preparation in the spin state that results in the bright (dark) distribution.

The error probabilities $P(1|\downarrow)$ and $P(0|\uparrow)$ are, thus,
\begin{align}
    &P(1|\downarrow) = \sum_{k=0}^{N_r-1} (1-\epsilon_0) p[k,\bar{n}_b] + \epsilon_0 p[k,\bar{n}_d], \\
    &P(0|\uparrow) = \sum_{k=N_r}^{\infty} \epsilon_1 p[k,\bar{n}_b] + (1-\epsilon_1) p[k,\bar{n}_d],
\end{align}
respectively. Evaluating using the statistics given by Eq.~\ref{eqn:poisson} yields
\begin{equation}
    F_r = \frac{1}{2} + \frac{f_0}{2}\left(\frac{\Gamma[N_r,\bar{n}_d]-\Gamma[N_r,\bar{n}_b]}{\Gamma[N_r,0]} \right),
    \label{eqn:Fr_general}
\end{equation}
where $\Gamma[N_r,n]=\int_n^\infty t^{N_r-1} e^{-t} dt$ is the Gamma function and $f_0=1-\epsilon_0-\epsilon_1$ sets the maximum attainable readout fidelity. This can be simplified if $N_r=1$, a choice which optimizes readout fidelity for all measurements in this work. Doing so yields
\begin{equation}
    F_r = \frac{1}{2} + \frac{f_0}{2} \left( e^{-\bar{n}_d} - e^{-\bar{n}_b} \right).
    \label{eqn:Fr_model}
\end{equation}

\subsection{Spin readout using cavity quantum electrodynamics}
\label{appendix:cqed_readout}
As discussed in the main text Sec.~\ref{sec:weak_meas}, comparison of emission to measurement-induced dephasing allows for a general way to characterize measurement efficiency, defined as the total loss between qubit and detector. In this section, we show how this idea is applicable to a strongly interacting cavity QED system, driven near resonance to interrogate the spin state. This style of spin readout has previously been demonstrated using SiV$^-$ centers in diamond \cite{nguyen:2019prl,nguyen:2019,kindem:2020,stas:2022} and is a natural next step for the SnV$^-$ platform.

To understand how spin readout works in a probed cavity QED system, consider a qubit whose state is described by density matrix $\hat{\rho}$ and which is prepared in the spin superposition $(\ket{\downarrow} + \ket{\uparrow})/\sqrt{2}$. This spin may interact in a spin-selective way with another quantum state such as the electromagnetic field in a driven cavity. Upon interaction, this other state can acquire a spin-dependent phase and/or amplitude shift, such that the joint quantum state of the spin and cavity is \cite{clerk:2010,blais:2021}
\begin{equation}
    \ket{\psi} = \frac{1}{\sqrt{2}}\left(\ket{\downarrow}\ket{\alpha_\downarrow} + \ket{\uparrow}\ket{\alpha_\uparrow} \right),
    \label{eqn:joint_state_for_readout}
\end{equation}
with density matrix $\hat{\sigma} = \ket{\psi}\bra{\psi}$. Here, $\ket{\alpha_\downarrow}$ and $\ket{\alpha_\uparrow}$ are the quantum states of light in the driven cavity, and are projectively measured to gain information about the spin.

\begin{figure}[htb!] 
\begin{center}
\includegraphics[width=1.0\columnwidth]{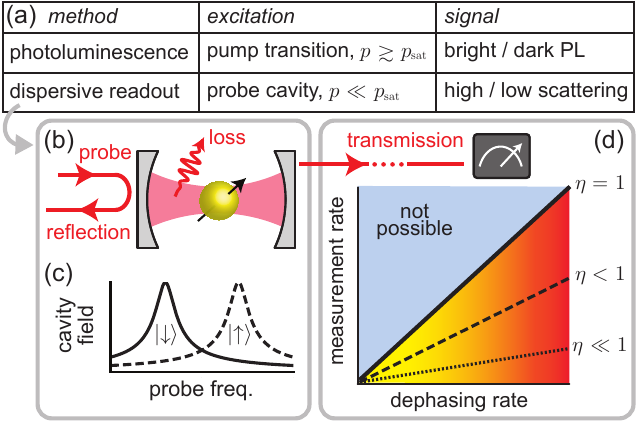}
\caption{(a) Spin readout discussed thus far in this work is based on the presence or absence of photoluminescence under a resonant optical pump. Alternatively, in a cavity quantum electrodynamics system (b) for which an SnV$^-$'s optical transition interacts with a cavity, the spin may be measured via spin-dependent scattering from a probe tone. (c) This technique, known as ``dispersive readout'' requires the cavity to have a spin-dependent dispersive shift $\chi$ that is appreciable compared to the cavity linewidth $\kappa$ \cite{nguyen:2019,nguyen:2019prl,blais:2021,antoniadis:2023}. (d) For both techniques, the measurement efficiency $\eta$ may be precisely determined from a comparison between measurement-induced dephasing and the rate at which information is acquired by a measurement apparatus \cite{clerk:2010}. The top left of this cartoon plot is inaccessible, as it would indicate the impossible case of a detector which gains information about the qubit state without sufficient collapse of its superposition, and $\eta=1$ bounds physically realizable measurements.}
\label{fig:cqed}
\end{center}
\end{figure}

\subsubsection{Measurement-induced dephasing of a cavity QED system}
The overlap between $\ket{\alpha_\downarrow}$ and $\ket{\alpha_\uparrow}$ sets both how much the spin state may be distinguished by measurement \cite{clerk:2010,blais:2021} and how much dephasing is caused by measurement.

To show this, following the discussion in Ref.~\cite{rosenthal:2021} consider a general quantum measurement described by the positive
operator-valued measure (POVM) $\hat{\Pi}_n = \hat{K}_n^\dag \hat{K}_n$. 
Here, $\hat{K}_n^\dag$ and $\hat{K}_n$ are Krauss operators that correspond to measurement of a light field in the number basis, and $n$ is a non-negative integer. After measurement, the quantum state of the joint spin-photonic system is described by density matrix $\hat{\sigma}' = \sum_n \hat{K}_n \hat{\sigma} \hat{K}_n^\dag$ \cite{kraus:1971}.

We model readout of the joint system $\hat{\sigma}$ using the POVM $\hat{\Pi}_n = \mathbb{I} \otimes \ket{n} \bra{n}$, which describes measurement of the second system in the basis $\ket{n}$. After a measurement outcome $n$, the joint system is:
\begin{equation}
    \hat{\sigma}_n^{'} = \frac{1}{2 \nu_n} \begin{pmatrix}
    |\braket{n|\alpha_\downarrow}|^2 &  \braket{n|\alpha_\downarrow} \braket{\alpha_\uparrow|n} \\
    \braket{n|\alpha_\uparrow} \braket{\alpha_\downarrow|n} & |\braket{n|\alpha_\uparrow}|^2
    \end{pmatrix} \otimes \ket{n} \bra{n},
\end{equation}
where $\nu_n = \mathrm{Tr}\left( \hat{\Pi}_n \hat{\sigma} \right)$ is a normalization factor.

Taking the partial trace $\hat{\rho}^{'} = \mathrm{Tr}_2\left( \sum_n \nu_n \hat{\sigma}_n^{'} \right)$ over the second system gives the spin's postmeasurement density matrix:
\begin{equation}
    \hat{\rho}^{'} = \frac{1}{2} \begin{pmatrix}
    1 &  \braket{\alpha_\uparrow|\alpha_\downarrow} \\
    \braket{\alpha_\downarrow|\alpha_\uparrow} & 1
    \end{pmatrix}.
    \label{eqn:post_measurement_spin_state}
\end{equation}
From Eq.~\ref{eqn:post_measurement_spin_state}, we see that spin coherence is reduced to:
\begin{equation}
    \hat{\rho}_{\downarrow \uparrow}^\prime = \frac{1}{2} \braket{\alpha_\downarrow|\alpha_\uparrow}.
    \label{eqn:post_measurement_coherence}
\end{equation}

For example, consider a cavity that is in one of two spin-dependent coherent states:
\begin{align}
    \ket{\alpha_\downarrow} = e^{-\frac{1}{2}|\alpha_\downarrow|^2} \sum_{n=0}^\infty \frac{\alpha_\downarrow^n}{\sqrt{n!}}\ket{n}, 
    \label{eqn:alpha_down} \\
    \ket{\alpha_\uparrow} = e^{-\frac{1}{2}|\alpha_\uparrow|^2} \sum_{n=0}^\infty \frac{\alpha_\uparrow^n}{\sqrt{n!}}\ket{n}.
    \label{eqn:alpha_up}
\end{align}
After the cavity state is projectively measured, the spin's coherence is
\begin{equation}
    |\hat{\rho}_{\downarrow \uparrow}^\prime| = \frac{1}{2}e^{-|\alpha_\downarrow - \alpha_\uparrow|^2/2}.    \label{eqn:post_measurement_coherence_explicit}
\end{equation}
Therefore, if $\ket{\alpha_\downarrow}$ and $\ket{\alpha_\uparrow}$ are identical, the spin is not measured and is not dephased. On the other hand, if $\ket{\alpha_\downarrow}$ and $\ket{\alpha_\uparrow}$ are orthogonal, then coherence is destroyed; this is the limit of projective measurement, in which maximal information may be acquired about the spin.

\subsubsection{Measurement of the spin state}
We now apply these general principles of quantum measurement to understand readout of a probed cavity QED system. Consider the model of an optical cavity coupled to a single ``atom'', Fig.~\ref{fig:cqed}. As in Fig.~\ref{fig:lambda}, this atom is treated as a ``lambda system'' with spin ground states $\ket{\downarrow}$ and $\ket{\uparrow}$, which have transitions A1 or A2 to an excited state $\ket{A}$, respectively. Coupling between the atom and cavity can change the cavity frequency by the dispersive shift $\pm \chi$, dependent on the spin state. The magnitude $\chi$ depends on the coupling between atom and cavity and their detuning is related to the cooperativity between them (desired to be high). A noticeable shift can occur, for instance, if the atom is tuned such that the A1 transition is resonant with the cavity, coupling between the A1 transition and cavity is strong compared to overall system loss, and the A2 transition is far off resonant (so coupling between this transition and the cavity is negligible). Such a system is well approximated by a limit of the Jaynes-Cummings Hamiltonian \cite{reiserer:2015} $\hat{H}/\hbar \approx (\omega_c + \chi \hat{\sigma}_z) \hat{a}^\dag \hat{a} + \omega_q \hat{\sigma}_z/2$, where $\hat{a}$ is the cavity field operator, $\hat{\sigma}_z$ is the Pauli operator describing the atom's spin state, $\omega_c$ is the bare cavity frequency, and $\omega_q$ is the spin transition frequency. Finally, readout requires nonunitary interaction with a measurement apparatus. The total loss rate of the cavity is described by the linewidth $\kappa$, which includes internal loss $\kappa_\mathrm{int}$ coming from intrinsic cavity loss of the loaded cavity, as well as external loss $\kappa_\mathrm{ext}$ due to external coupling.

Under a near-resonant drive we can use input-output theory to solve for the steady-state field inside the cavity. Driven far below saturation and in the limit where $\gamma \ll \kappa$, this becomes \cite{blais:2021}
\begin{equation}
    \alpha_{\downarrow,\uparrow}(\delta) = \frac{\epsilon}{\sqrt{(\kappa/2)^2 + (\delta\pm\chi)^2}},
    \label{eqn:field_amp}
\end{equation} 
where $\delta$ is the detuning of the drive from the bare cavity frequency, the drive rate $\epsilon$ is proportional to the drive amplitude, and $\epsilon^2$ is proportional to drive power. (Relaxing the assumption of $\gamma\ll\kappa$ complicates Eq.~\ref{eqn:field_amp}, but retains this proportionality \cite{walls:1994}.) The response of Eq.~\ref{eqn:field_amp} is plotted in Fig.~\ref{fig:cqed}b for a choice of $\chi=2\kappa$.

Readout works by inferring the spin state based on a measurement of $\ket{\alpha_\downarrow}$ or $\ket{\alpha_\uparrow}$. Consider a steady-state system where the cavity is driven for a readout window $\tau$ much longer than the transient dynamics of the field's ring-up or -down time, such that $\tau \gg 1/\kappa$. The mean number of expected photon counts at a detector is $\bar{n}_b = \eta \kappa \tau |\alpha_\downarrow|^2$ if the cavity is in $\ket{\alpha_\downarrow}$ and $\bar{n}_d = \eta \kappa \tau |\alpha_\uparrow|^2$ if the cavity is in $\ket{\alpha_\uparrow}$. Here, $\kappa$ sets the overall rate at which photons leave the cavity and are, thus, incoherently lost. The measurement efficiency $\eta$ is again the overall fraction of these photons that are actually detected.

Notably, unlike the photoluminescence-based readout discussed elsewhere, here $\eta$ does not suffer from loss due to the emitter's quantum efficiency or fraction of emission into the phonon sideband vs zero-phonon line. This is because we consider the limit of low enough excitation power ($p \ll p_\mathrm{sat}$), such that the scattered probe dominates signal and photoluminescence is negligible. Cyclicity becomes important only as probe power is increased so that excitation or photoluminescence appears as a concurrent effect.

In analogy with Eq.~\ref{eqn:emission_model} in the main text, we model the difference in collected counts between the bright and dark distributions to be $\bar{n}_b - \bar{n}_d = \eta \kappa \tau (|\alpha_\downarrow|^2 - |\alpha_\uparrow|^2)$. We would, therefore, experimentally measure
\begin{equation}
    \bar{n}_b - \bar{n}_d = \mathcal{A} p, 
    \label{eqn:emission_vs_power_probe_readout} 
\end{equation}
where $\mathcal{A} = \eta \kappa \tau \left(|\alpha_\downarrow|^2 - |\alpha_\uparrow|^2 \right) / p$ is a constant of proportionality that can be fit to experimental data in analogy to Fig.~\ref{fig:power}b. Notice that $\mathcal{A}$ is independent of drive power, because the intracavity photon number (either $|\alpha_\downarrow|^2$ or $|\alpha_\uparrow|^2$) scales as $\epsilon^2$, which is proportional to $p$.

In summary, the difference between the measured distributions associated with the spin being in $\ket{\uparrow}$ or $\ket{\downarrow}$ allows for determination of the spin state. These distributions grow further apart as integration time increases, drive power increases, and measurement efficiency increases. The limit where these distributions are sufficiently separated allows for a high-fidelity single-shot measurement. In practice, it is desirable to have a short integration time for faster readout and a low drive amplitude to avoid other dynamics, and so it remains important to have high efficiency.

\subsubsection{Determining efficiency of a cavity QED system}
We have seen that measurement of the probe both dephases the qubit and allows for determination of the spin state. As for the case of readout using photoluminescent emission (e.g. Eq.~\ref{eqn:eta_compare_rates}), we compare these effects to quantify the measurement efficiency $\eta$. 

After projective measurement of a spin-dependent probe, the spin's postmeasurement coherence decreases by Eq.~\ref{eqn:post_measurement_coherence_explicit}. In a steady-state cavity that is continuously driven for duration $\tau$, the field in the cavity is continuously being lost and/or measured at rate $\kappa$, and continuously repopulated by the drive. We can understand the overall dephasing given by Eq.~\ref{eqn:post_measurement_coherence_explicit} to occur at some dephasing rate $\Gamma_\phi$, such that the spin's phase coherence decays in magnitude by a factor of $e^{-\Gamma_\phi \tau}$ over time $\tau$. The dephasing rate
\begin{equation}
    \Gamma_\phi = \frac{1}{2} \left| \alpha_\downarrow - \alpha_\uparrow \right|^2 \kappa
    \label{eqn:dephasing_probe_readout}
\end{equation}
is the overlap of the probe states times the loss rate of the cavity.

Like the collected signal in Eq.~\ref{eqn:emission_vs_power_probe_readout}, the dephasing rate is proportional to drive power such that
\begin{equation}
    \Gamma_\phi = \mathcal{B} p,
    \label{eqn:dephasing_vs_power_probe_readout}
\end{equation}
where $\mathcal{B} = \kappa \left| \alpha_\downarrow - \alpha_\uparrow \right|^2 / (2p)$ is a proportionality constant that may be experimentally determined by the method described in Fig.~\ref{fig:dephasing}. (Note that $\mathcal{B}$ is constant with power, because $\left| \alpha_\downarrow - \alpha_\uparrow \right|^2 \propto p$.)

In analogy with Eq.~\ref{eqn:eta_compare_rates}, comparison between constants $\mathcal{A}$ and $\mathcal{B}$ yields the measurement efficiency
\begin{equation}
    \eta = \frac{\mathcal{A}}{2 \mathcal{B} \tau} \times f.
\end{equation}
Here, the added factor $f = |\alpha_\downarrow - \alpha_\uparrow|^2 / (|\alpha_\downarrow|^2 - |\alpha_\uparrow|^2)$ depends on the spin-dependent steady-state cavity field, which is a function of $\chi$ and $\kappa$. These parameters may experimentally determined by spectroscopic measurements of the cavity in order to determine $f$. In the limit $\kappa \ll \chi$, this simplifies further to $f \approx 1$ and $\eta \approx \mathcal{A} / (2 \mathcal{B} \tau)$.

In summary, we have derived a general method to characterize the measurement efficiency of $\eta$ of a probed cavity QED system. Readout using this method relies on scattering of a probe tone off a dispersively shifted cavity, such that the spin state is inferred by measurement of the scattered field. While this method is fundamentally different than the spin readout via detection of photoluminescent emission discussed elsewhere in this work, both techniques are well described by the general theory of quantum measurement including measurement-induced dephasing. We emphasize that this section has direct analogy to well-established techniques in the readout of superconducting qubits in circuit QED architectures \cite{clerk:2010,bultink:2018,blais:2021,rosenthal:2021,lecocq:2021,white:2023}.

Finally, we emphasize the applicability of this section to readout of color center spin qubits integrated into cavity QED systems, including recent experiments using SiV$^-$ centers \cite{nguyen:2019,nguyen:2019prl,bhaskar:2020,stas:2022} or quantum dots \cite{antoniadis:2023}. In contrast to the SiV$^-$ (737 nm zero-phonon line), the shorter emission wavelength of the SnV$^-$ (620 nm) can present a technical challenge to nanofabrication, but dispersive readout of the SnV$^-$'s spin state should work comparably well so long as similar cooperativity can be reached. Recent demonstrations of Purcell enhancement of an SnV$^-$ center integrated into a nanophotonic cavity suggest that the SnV$^-$ platform is close to this regime \cite{rugar:2021,kuruma:2021,herrmann:2023}, and, thus, disperive readout of an SnV$^-$ spin is a promising avenue for future experiments.
\section{Extended discussion}

\subsection{Understanding the S\MakeLowercase{n}V$^-$ Hamiltonian}
\label{sec:fit_optical_transitions}
The SnV$^-$ Hamiltonian is predicted by \textit{ab initio} calculations in Ref.~\cite{thiering:2018}. In our previous work, Ref.~\cite{rosenthal:2023}, these calculations are compared to experimental values obtained by fitting the eigenvalues of the SnV$^-$ Hamiltonian to measurement of its level structure vs magnetic field orientation. There are many parameters in this fit including strain, spin-orbit coupling, the orbital Zeeman effect, anisotropy of the Zeeman effect, and the vector orientation of the spin dipole. This fit is underdetermined if every parameter of the SnV$^-$ Hamiltonian is taken as a free parameter, and so to constrain the fit we made assumptions: (1) The direction normal to the surface of the chip is parallel to the $\hat{z}$ axis of the magnetic coil. (2) The magnet calibration is accurate and magnet hysteresis is negligible. (3) Higher-order terms in the SnV$^-$ Hamiltonian are neglected \cite{thiering:2018}.

The values of the SnV$^-$ Hamiltonian reported in Table~1 in Ref.~\cite{rosenthal:2023} rely on these assumptions, which are imperfect at some level. While this analysis was a more precise determination of the SnV$^-$ Hamiltonian than prior experimental work \cite{iwasaki:2017,rugar:2019,gorlitz:2020,trusheim:2020}, it can be improved upon by future experiments that do not rely on these assumptions. For example, subsequent work in Ref.~\cite{karapatzakis:2024} reports slightly different values of the SnV$^-$ Hamiltonian but constrains the fit by studying multiple emitters, by using a three-axis vector magnet to optimize alignment with the spin dipole, and by quantifying the magnet's calibration.

\subsection{Estimating angular misalignment}

Understanding the SnV$^-$ Hamiltonian helps to understand spin polarization and readout, including by determining alignment between the magnetic field and spin dipole moment. In our current experiment we use a two-axis vector magnet, only, which prohibits arbitrary alignment between the magnet and spin. To estimate the minimum misalignment we compare the measurement of cylicity vs $\zeta$ (Fig.~\ref{fig:intro}e) to a model based on the driven SnV$^-$ Hamiltonian (see Appendix B.3 in Ref.~\cite{rosenthal:2023}) and using the Hamiltonian parameters from Ref.~\cite{karapatzakis:2024}. This model closely follows our data for an azimuthal misalignment of $\varphi \approx 10\degree$. Use of the alternate Hamiltonian parameters reported in Ref.~\cite{rosenthal:2023} has minimal effect on this result.

\subsection{Importance of strain and magnetic field alignment}
\begin{figure}[htb!] 
\begin{center}
\includegraphics[width=1.0\columnwidth]{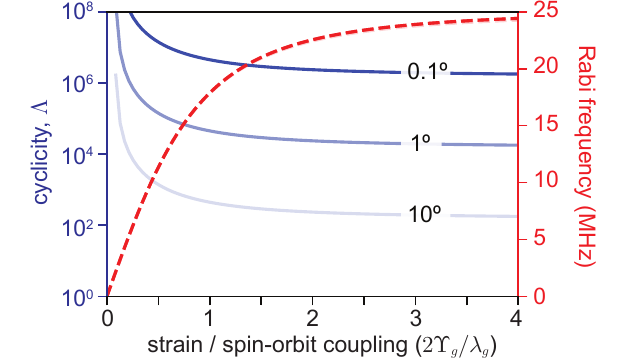}
\caption {Strain considerations for cyclicity and spin control. \textit{X} axis: the ratio of the variable ground-state strain  $2\Upsilon_g$, to the fixed ground-state spin-orbit coupling $\lambda_g/2\pi = 822 \units{GHz}$ \cite{karapatzakis:2024}. Strain $2\Upsilon_g$ is defined such that the ground-state splitting is $\sqrt{\lambda_g^2 + (2\Upsilon_g)^2}$. Left \textit{y} axis (blue): cyclicity (solid lines) for dc magnetic field misalignment of $0.1\degree$, $1\degree$, and $10\degree$. Right \textit{y} -axis (red): Rabi frequency when driven by an ac microwave field perpendicular to the spin dipole moment (dashed lines). Results fall on top of each other for field misalignment of $0.1\degree$, $1\degree$, and $10\degree$.}
\label{fig:tradeoff}
\end{center}
\end{figure}

Both magnetic field misalignment and strain have an important impact on cyclicity and microwave Rabi frequency and, thus, on the SnV$^-$'s utility as a spin-photon interface. In Fig.~\ref{fig:tradeoff}, we plot a numerical simulation of these effects based on the SnV$^-$ Hamiltonian model discussed in Ref.~\cite{karapatzakis:2024}, using an excited state strain of $2\Upsilon_e/2\pi=268~\unit{GHz}$ drawn from Ref.~\cite{rosenthal:2023} (the same SnV$^-$ center used in this work). Results do not qualitatively change if instead we use other Hamiltonian parameters from Ref.~\cite{rosenthal:2023}.

In this simulation, we sweep ground-state strain, parameterized as a fraction of the spin-orbit coupling $\lambda_g/2\pi = 822~\unit{GHz}$ \cite{karapatzakis:2024}. In the low strain limit $2\Upsilon_g \ll \lambda_g$, cyclicity diverges, while the spin transition dipole becomes forbidden and the microwave Rabi frequency goes to zero. In the limit of $2\Upsilon_g \gg \lambda_g$, cyclicity is lowest, but the microwave Rabi rate plateaus to that of a free electron. The SnV$^-$ used this work has a strain of $2 \Upsilon_g / \lambda_g \approx 0.43$ \cite{rosenthal:2023}, which results in a combination of cyclicity of order $10^3-10^4$ and a microwave Rabi frequency of order $1-10 \units{MHz}$ (Fig.~\ref{fig:intro}f,g). Note that these simulations use an ac microwave drive field perpendicular to the spin dipole $\Vec{\mu}$, along with a slight angular misalignment of the dc magnetic field $|\Vec{B}| = 125~\unit{mT}$. Simulations in Fig.~\ref{fig:tradeoff} depend on many parameters, including that of the SnV$^-$ Hamiltonian, which is still a subject of active research \cite{thiering:2018,iwasaki:2017,rugar:2019,trusheim:2020,gorlitz:2020,rosenthal:2023,karapatzakis:2024}, and strain in the excited state manifold, which is not independent of strain in the ground-state manifold but here is held constant for simplicity. Thus, we take the simulations in Fig.~\ref{fig:tradeoff} as a qualitative example of the physics at work in group-IV centers, not as a precise quantitative prediction; see Refs.~\cite{rosenthal:2023,pieplow:2024} for further quantitative discussion.

The simulations in Fig.~\ref{fig:tradeoff} illustrate a strain-dependent relationship between spin control and readout. Decreased strain increases cyclicity, which improves readout to the detriment of spin control. Increased strain improves spin control but to the detriment of lower cyclicity for a misaligned magnetic field. Cyclicity increases as the magnetic field becomes more aligned; this slows the spin polarization time via resonant driving (e.g., Fig.~\ref{fig:intro}d) but is helpful for spin readout. Since field alignment has negligible effect of microwave Rabi frequency, it is desirable to work in a highly aligned field, in parallel to previous SiV$^-$ experiments \cite{sukachev:2017}.

Finally, we emphasize that the interplay between strain and spin-orbit coupling discussed here is broadly applicable to all group-IV centers. Of group IV's, recent SiV$^-$ experiments motivated by quantum networking operate in the strained $2\Upsilon_g/\lambda_g > 1$ limit which is helpful for spin control and which increases ground-state splitting, mitigating temperature-related decoherence \cite{sukachev:2017,nguyen:2019,nguyen:2019prl,bhaskar:2020,stas:2022}. The SiV$^-$'s spin-orbit coupling term of only approximately $50~\unit{GHz}$ makes it easier to reach this limit than for the SnV$^-$, whose spin-orbit coupling is approximately $820~\unit{GHz}$, which constrains recent work to $2\Upsilon_g/\lambda_g \lesssim 1$ \cite{rosenthal:2023,guo:2023,karapatzakis:2024}. Our results here favorably show that single-shot readout and rapid spin control are compatible for the SnV$^-$ even for the moderate strain of $2\Upsilon_g/\lambda_g \approx 0.43$, suggesting a wide regime of parameter space over which SnV$^-$-based devices can operate as a spin-photon interface with application to, for example, quantum networks.

\subsection{Comparison to prior work} 

\begin{table*}[htb!]
\caption{Comparison to selected prior work on single-shot readout of color centers in diamond.}
  \begin{center}
    \begin{tabular}{ |p{3cm}|p{2.25cm}|p{1.75cm}|p{1.75cm}|p{1.75cm}|p{1.75cm}|p{2.25cm}|  }
         \hline
         Reference & Qubit & Fidelity, $F_r$ & Counts, $\bar{n}_b$ & Duration, $\tau$ & Device & Spin control? \\
         \hline
         \hline
          Robledo \textit{et al.} \cite{robledo:2011} & NV$^-$ & 93\% & 6.4 & $40 \units{\mu s}$ & SIL & Yes \\
         \hline
          Sukachev \textit{et al.} \cite{sukachev:2017} & SiV$^-$ & 89\% & 6.2 & $20 \units{ms}$ & Bulk & Yes \\
         \hline
          G\"orlitz \textit{et al.} \cite{gorlitz:2022} & SnV$^-$ & 74\% & 1.2 & $200 \units{\mu s}$ & Bulk & No \\
          \hline
          Parker \textit{et al.} \cite{parker:2024} & SnV$^-$ (nuclear) & 80\% & 1.6 & $15 \units{\mu s}$ & Waveguide & No \\          
         \hline
          This work & SnV$^-$ & 87.4\% & 4.3 & $50 \units{\mu s}$ & Mesa & Yes \\
         \hline            
    \end{tabular}
  \label{tab:comparison_to_prior_work}
  \end{center}
\end{table*}

Despite subtleties in determining the SnV$^-$ Hamiltonian and angular misalignment, it is clear that the readout fidelity reported in this work uses an SnV$^-$ center which is only somewhat aligned with the external $\Vec{B}$ field, has a spin polarization time of approximately $20 \units{\mu s}$, and has a cyclicity of approximately $2\times10^3$. At this operating condition, we achieve single-shot readout of the SnV$^-$ center's electronic spin with an average of up to approximately $4$ photons collected during a $50 \units{\mu s}$ integration window. In Table~\ref{tab:comparison_to_prior_work}, we compare these results to prior examples of single-shot spin readout of color centers in diamond.

Looking to the future, the misalignment and modest collection efficiency of our experiment emphasize how much SnV$^-$ readout can be improved. Even with these nonidealities, the rate of photon collection in our work is much greater than that in the first demonstration of single-shot readout of an SiV$^-$ electron spin \cite{sukachev:2017}, which reported an average of approximately $6$ photons collected during a 20 ms readout window. This may reflect the SnV$^-$'s higher quantum efficiency $\eta_q$: Ref.~\cite{thiering:2018} calculates $\eta_q=0.14$ for the SiV$^-$ and $\eta_q=0.91$ for the SnV$^-$. Subsequent work on SiV$^-$ readout has shown greater collection efficiency over faster timescales by optimized use of nanophotonics \cite{nguyen:2019prl,nguyen:2019,bhaskar:2020,stas:2022}. Our results here suggest such improvement will also be possible using the SnV$^-$.

We also compare our results to Ref.~\cite{gorlitz:2022}, to our knowledge the only prior published single-shot readout of an SnV$^-$ electronic spin. Reference~\cite{gorlitz:2022} demonstrates readout of $74\%$ fidelity (max of $1.2$ mean counts per shot) collected during a $200 \units{\mu s}$ readout window. This result is achieved in a bulk sample on an unstrained SnV$^-$. Lack of strain yields high cyclicity with reduced sensitivity to alignment of the magnetic field but also suppresses the magnetic dipole transition necessary for spin control \cite{rosenthal:2023}. Finally, we note that Ref.~\cite{parker:2024} measures single-shot readout of a nuclear spin via a waveguide-integrated SnV$^-$, obtaining a single-shot nuclear spin readout fidelity of $80\%$ in $15 \units{\mu s}$. In comparison, our work demonstrates a regime in which \textit{both} single-shot readout of an SnV$^-$'s electronic spin and rapid microwave control of this spin are possible.

Finally, readout fidelity can be boosted by performing spin-to-charge conversion \cite{anderson:2022}. We have already demonstrated charge state readout with our CRC check. A spin-dependent ionization process could be achieved with a narrow line, spin-selective laser and a high power ionization laser. Combined, they could provide increased photons measured per shot, separating the bright and dark histograms and boosting fidelity.

\section{Experimental details}

\begin{figure*}[htb!] 
\begin{center}
\includegraphics[width=2.0\columnwidth]{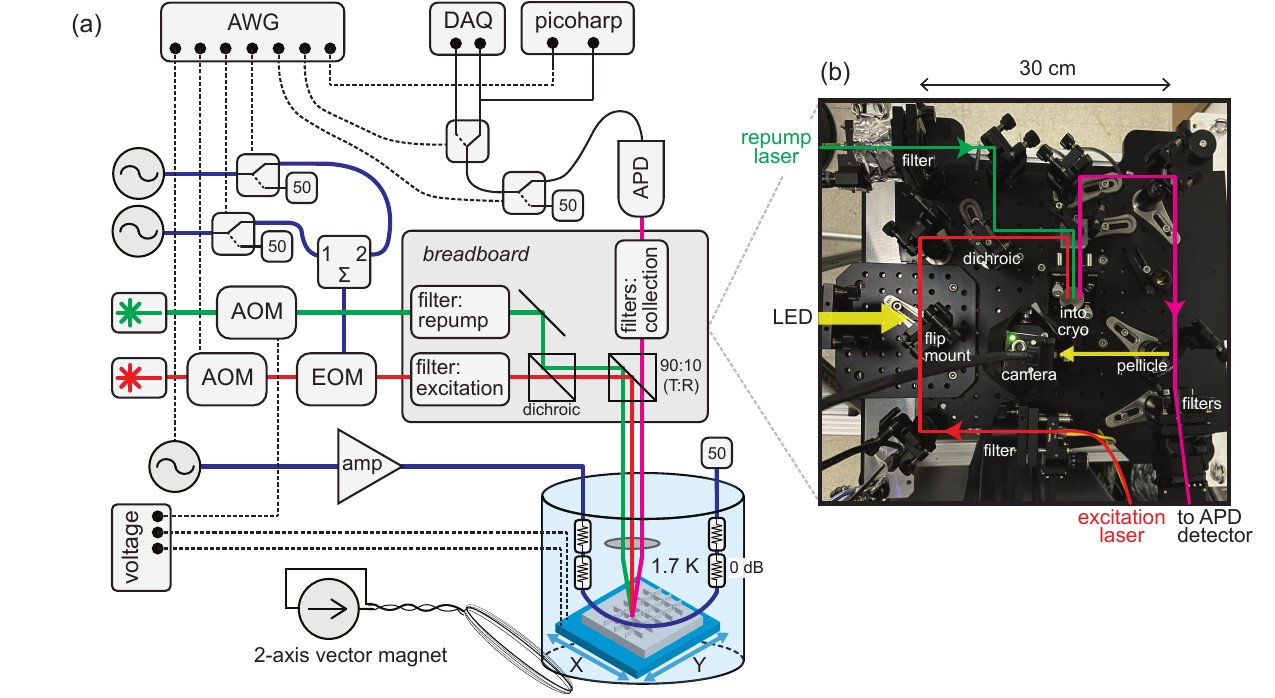}
\caption{
(a) Experimental schematic. (b) Photograph of the optical breadboard.
}
\label{fig:schematic}
\end{center}
\end{figure*}

\subsection{Setup}
\label{sec:setup}
The setup used in this experiment, Fig.~\ref{fig:schematic}, is similar to that in Ref.~\cite{rosenthal:2023} but optimized for higher transmission on the collection path. The repump (532 nm) and resonant (M-Squared) lasers are both modulated by acousto-optic modulators (4C2C-532-AOM and 4C2C-633-AOM, Gooch and Housego). Typically, we operate with the excitation laser detuned by several GHz from the SnV$^-$'s optical transitions and drive these transitions using the first sideband created by an electro-optic modulator (PM-0S5-PFU-PFU-620, Eospace). By combining two microwave signal generators (SG396, SRS) using a microwave power combiner, we can drive the A1 transition only, the B2 transition only, or both simultaneously as is necessary for implementing charge resonance checks. Pulsing is controlled by an arbitrary waveform generator (Pulse Streamer 8/2, Swabian).

For spin control, microwaves are delivered by a wire bond draped over the diamond chip as in Ref.~\cite{rosenthal:2023}, and using a similar microwave power ($48 \units{dBm}$ into the cryostat). The wire bond has been replaced from that used in Ref.~\cite{rosenthal:2023}; we attribute the slower Rabi frequency here to the new wire bond simply being further from the spin.

To focus excitation light and collect signal confocally, a cryogenic objective (LT-APO/VISIR/0.82, Attocube) is placed directly above the chip. This objective has a focal length of 2.87 mm, a numerical aperture of 0.82, and a working distance of 0.65 mm. Signal is then routed out of the cryostat to a 90:10 beam splitter, whose 90\% port leads to a free space to fiber coupler. This coupler is connected by a multimode optical fiber (FG025LJA, core diameter $25\pm3\units{\mu m}$, Thorlabs) to a single-photon counting module (SPCM-AQR-15-FC, Perkins Elmer). Detection events are time tagged (Picoharp300, PicoQuant).

Before the fiber coupler, we place a 635 nm long pass filter (635 nm Longwave EdgeBasic, Semrock) to collect the SnV$^-$'s phonon sideband (PSB) only, thus filtering most excitation scatter and also the SnV$^-$'s zero-phonon line. Note that this wavelength still filters a fraction of the PSB; a different filter closer to the zero-phonon line would have lead to greater counts, e.g., a filter at 625 nm. Finally, a 532 nm notch filter (532 nm StopLine, Semrock) is also placed on the collection path to mitigate noise from the repump laser; even though this laser is toggled off during readout, there is some nonzero feedthrough of the AOM used to toggle the green laser which can scatter into the collection path, and so this notch filter is still useful.

\subsection{Sources of loss}
\label{sec:loss_discussion}

\begin{table*}[htb]
\caption{Estimated sources of loss.}
  \begin{center}
    \begin{tabular}{ |p{4.25cm}|p{2cm}|p{4.25cm}|p{6.0cm}|  }
         \hline
         Source & Efficiency & Reference or method & Path to improve \\
         \hline
         \hline
         Quantum efficiency $\eta_q$ & $80\%-90\%$ & \cite{iwasaki:2017,thiering:2018} & \\
         \hline
         Debye-Waller factor $\eta_\mathrm{psb}$ & $43\%$ & \cite{gorlitz:2020} & Resonant collection (cross-polarization)  \\      
         \hline
         Scattering from chip $\eta_\mathrm{phot}$ & $5\%$ & Simulation, Fig.~\ref{fig:scattering} &  Nanopillars, waveguides, cavity-QED  \\   
         \hline
         Optical path loss $\eta_\mathrm{path}$ & $35\%$ & Transmission, at 670 nm & Optimized optics  \\               
         \hline
         Detector $\eta_\mathrm{det}$ & $65\%$ & Datasheet spec., at 650 nm & Superconducting detectors  \\               
         \hline
         Total $\eta$ & $0.4\%$ & $\eta = \eta_q \eta_\mathrm{psb} \eta_\mathrm{phot} \eta_\mathrm{path} \eta_\mathrm{det}$ & All of the above \\               
         \hline
    \end{tabular}
  \label{tab:loss_budget}
  \end{center}
\end{table*}

\begin{figure}[htb!] 
\begin{center}
\includegraphics[width=1.0\columnwidth]{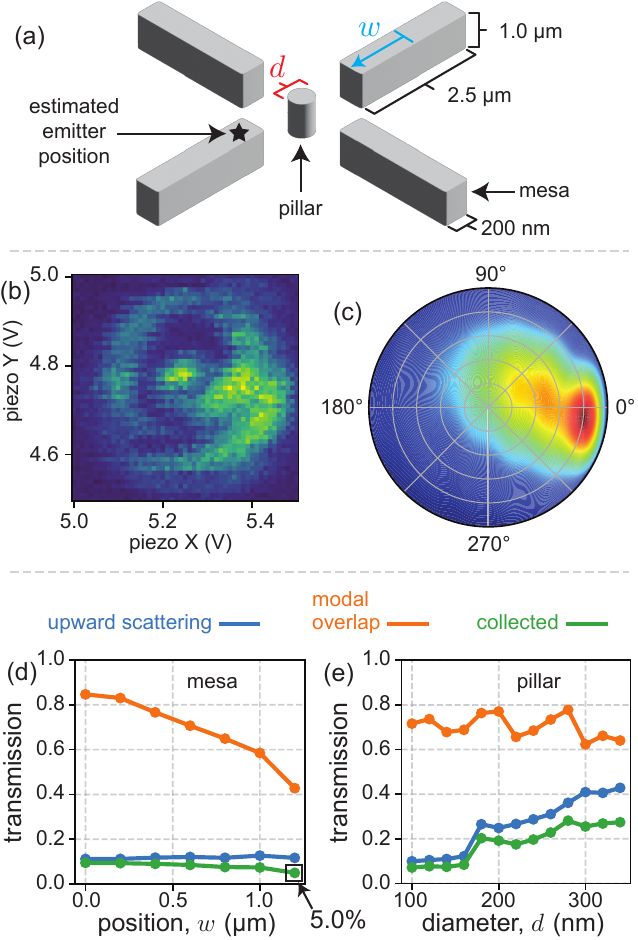}
\caption{
(a) Device. The $\langle100\rangle$ diamond chip contains arrays of nanophotonic structures, both pillars and mesas. Structures have a height of $1 \units{\mu m}$, and Sn atoms are implanted at a depth of approximately $90 \pm 20 \units{nm}$ \cite{rosenthal:2023}. (b) Scan to optimize focal position. The center of the measured lobe (which originates from imaging with a multimode fiber under slightly defocused conditions) corresponds to focal alignment at the edge of a mesa. A scan range of 0.2 V corresponds to a lateral distance of approximately $1 \units{\mu m}$. (c) Numerical simulation of the far-field scattering pattern from an SnV$^-$ center in a mesa. (d) Numerical simulation of 645 nm light collected from both a mesa and (e) a pillar. Simulations include the fraction of light scattered upward, the modal overlap of the objective and the scattering pattern, and their product, which gives the overall fraction of light routed to the collection path. From the simulation in (d), we conclude $\eta_\mathrm{phot} \approx 5\%$ of emission enters the collection path. In comparison, similar simulations show a scattering efficiency of $3\%$ from an emitter in bulk diamond at a depth of 100 nm.
}
\label{fig:scattering}
\end{center}
\end{figure}

In Table~\ref{tab:loss_budget}, we estimate the sources of inefficiency and optical loss expected for this work. Multiplying all sources of inefficiency yields $\eta = \eta_q \eta_\mathrm{psb} \eta_\mathrm{phot} \eta_\mathrm{col} \eta_\mathrm{det} \approx 0.4\%$. This qualitatively matches the measured values of $\eta \approx 0.2\%$ (Fig.~\ref{fig:readout}) and $\eta \approx 0.1\%$ (Fig.~\ref{fig:power}).

The dominant source of loss is internal reflection of light into the diamond sample, Fig.~\ref{fig:scattering}. By fine-tuning the emitter position using piezoelectric nanopositioners, we find this emitter to actually be in a mesa structure, rather than a nanopillar as stated in Ref.~\cite{rosenthal:2023}. For an SnV$^-$ center implanted at a depth of 100 nm and for structures $1 \units{\mu m}$ in height, simulations determine scattering from bulk diamond to be approximately $3\%$, scattering from a mesa is approximately $5\%-9\%$, Fig.~\ref{fig:scattering}d, and scattering from a pillar is approximately $7\%-28\%$, Fig.~\ref{fig:scattering}e. These values are reported at 645 nm, near the maximum of the SnV$^-$'s phonon sideband (PSB). Given the proximity of the color center to a pillar, we use values for an SnV$^-$ at the edge of the mesa structure and estimate the (average across the entire PSB) fraction $\eta_\mathrm{phot} \approx 5\%$ of emitted light is scattered into our collection path. This number accounts for the modal mismatch between the far field for the SnV$^-$ emission and the Gaussian mode accepted by the cryo-objective.

Other sources of loss include the emitter's quantum efficiency (the probability of nonradiative decay), filtering of the SnV$^-$'s ZPL, loss in the optical path, and detector inefficiency. Finally, Table~\ref{tab:loss_budget} may not capture every source of loss. Note that we use a 635 nm long pass filter to collect the PSB (see Appendix~\ref{sec:setup} for details); this removes approximately $14.5\%$ of the PSB light, and a cutoff of approximately $624 \units{nm}$ would be best to avoid extra loss \cite{iwasaki:2017,rugar:2019,trusheim:2020,gorlitz:2020}. Additionally, while we use multimode fiber to maximize photon collection, some imperfection in the coupling could still be present.

Optimized diamond nanophotonics provides a clear path toward high improvement in collection efficiency (e.g. approximately $25\%$ for inverse-designed grating couplers \cite{dory:2019}, and approximately $90\%$ for tapered fibers combined with 1D photonic crystal cavities \cite{burek:2017}). These devices have also been demonstrated in conjunction with stable SnV$^-$'s \cite{rugar:2020,rugar:2021,kuruma:2021,parker:2024,herrmann:2023,pasini:2023}. The other sources of loss can be improved upon too. For example, loss associated with filtering of the resonant excitation laser can be avoided by filtering by polarization rather than by wavelength (routinely done in the form of cross-polarization detection). Optical loss in the collection path can be mitigated with higher-quality optics, e.g., by using higher-reflectivity mirrors, and by replacing the 90:10 beam splitter on the collection path with a more transmissive component. Detector inefficiency can be improved by switching to state-of-the-art superconducting detectors. 

With improved nanophotonics only, we expect a collection efficiency of order $\times10$ or more, and with all possible improvements we expect an efficiency of order $\times100$ or more can be achieved in future experiments. Even modest improvements in efficiency will allow for much higher single-shot readout fidelity (since fidelity is non-linear with number of photons), operation at lower drive power (see Fig.~\ref{fig:power}c), faster readout, and single-shot readout at external magnetic fields with greater misalignment.

\subsection{Extended data}

\begin{figure*}[htb!] 
\begin{center}
\includegraphics[width=2.0\columnwidth]{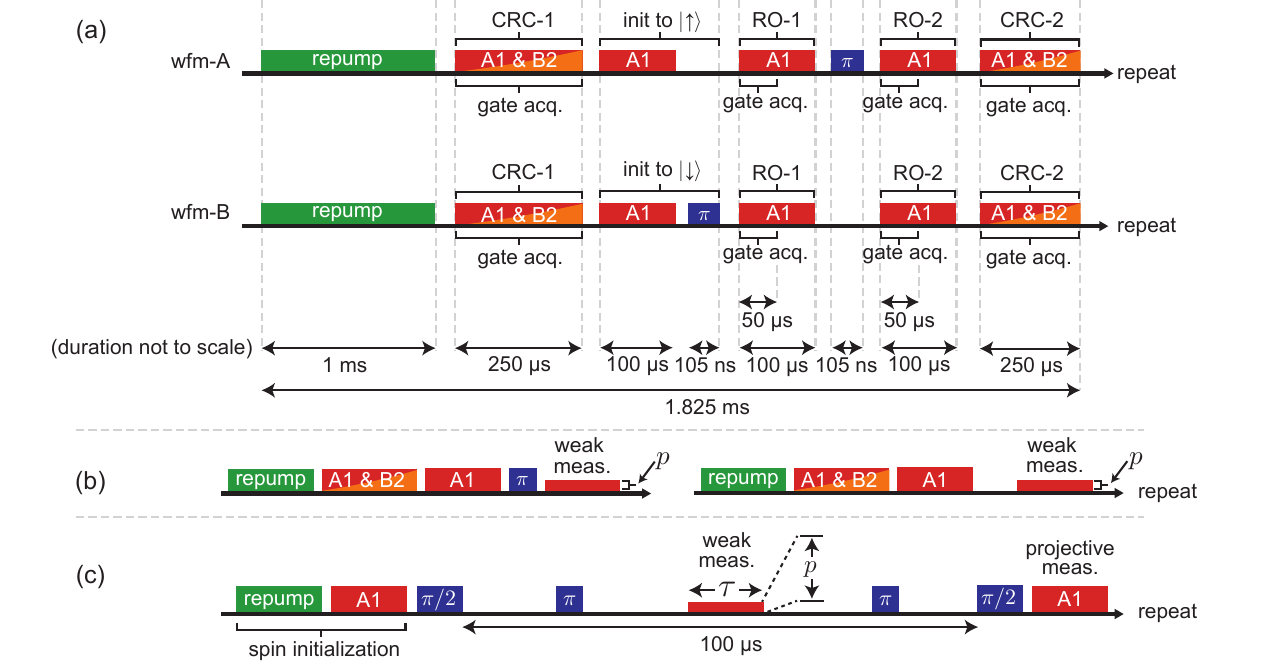}
\caption{
Timing diagrams of the experiments in: (a) Fig.~\ref{fig:readout}, (b) Fig.~\ref{fig:power}b, and (c) Fig.~\ref{fig:dephasing}. Note that (b) is a simplification because the waveform used is identical in timing to (a) including two readout steps and two CRCs; only the first of each is considered for Fig.~\ref{fig:power}b. Illustrations are not to scale. Steps are generally separated by $2 \units{\mu s}$ pauses (not shown). Weak measurement pulses in (b) and (c) resonantly drive the A1 transition, as in the spin polarization and projective measurement steps. 
}
\label{fig:readout_timing}
\end{center}
\end{figure*}

\subsubsection{Charge resonance checks}
\label{sec:CRC}

The readout characterization experiment in Fig.~\ref{fig:readout} includes charge resonance checks (CRCs) applied at both the beginning and end of each characterization cycle. A CRC consists of driving both the A1 and B2 transitions while simultaneously collecting emission. Many photons are collected during the CRC if the excitation is indeed resonant (e.g., neither the laser nor transition frequencies have drifted) and the system has not left the qubit subspace (e.g., it has not ``blinked off''). 

\begin{figure}[htb!] 
\begin{center}
\includegraphics[width=1.0\columnwidth]{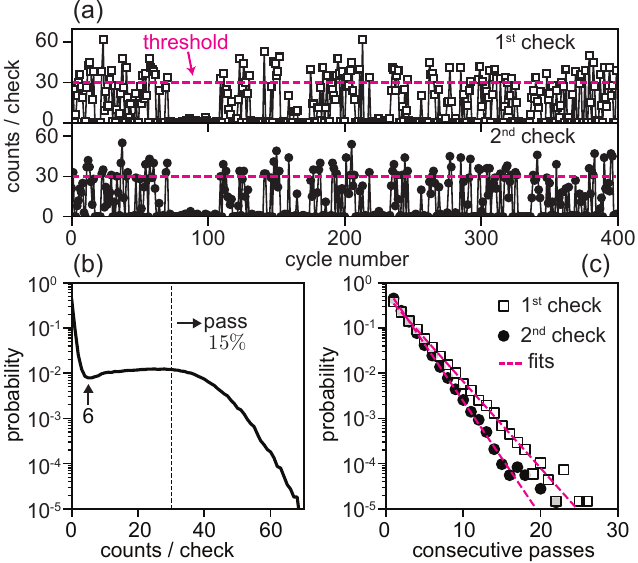}
\caption{
(a) Example of counts recorded during 400 CRCs. Each CRC consists of simultaneously driving the A1/B2 transitions. A threshold, here specified by the pink line at $N_c=30$, is used to discriminate between passing and failing checks. Data in the top (bottom) are from the first (second) check of the sequence used in Fig.~\ref{fig:readout} (see Fig.~\ref{fig:readout_timing}a for details), which is repeated $4.27\times10^5$ times. (b) Distribution of counts collected during all CRCs in the Fig.~\ref{fig:readout} dataset. Separate distributions with few or many counts per check are separated by an inflection point around six counts. Choosing a pass threshold of $N_c=30$ leads to approximately $15\%$ cycles passing the first check, and, of these, approximately $50\%$ pass the second check. (c) Probability of consecutive passes before a fail is recorded, also using the data from Fig.~\ref{fig:readout} and using $N_c=6$. Data are fit to exponential decay. Faster decay is observed for the second check, perhaps due to drive-induced charging in the intervening readout steps. In all panels, only data from wfm-A is plotted; data for wfm-B is nearly identical.
}
\label{fig:crc}
\end{center}
\end{figure}

The distribution of counts collected during a CRC are shown in Fig.~\ref{fig:crc}, using data from the ``wfm-A'' experiment in Fig.~\ref{fig:readout}. Data from wfm-B are nearly identical. The CRC result is recorded as a ``pass'' if $\ge N_c$ counts are measured and a ``fail'' if $< N_c$ counts are measured, where $N_c$ is the CRC discrimination threshold. Choosing $N_c=30$ for both CRCs for the data in Fig.~\ref{fig:readout}, we find that $7.4\%$ of cycles of wfm-A pass both CRCs, and $7.7\%$ of cycles for wfm-B pass both CRCs, out of $4.27\times10^5$ cycles total for each waveform. 

The probability that consecutive cycles pass either the first or second check is plotted in Fig.~\ref{fig:crc}c, using a pass threshold of $N_c=6$ (the inflection threshold between the dark and bright distributions, Fig.~\ref{fig:crc}b) and using a cycle time of $\tau_\mathrm{cycle} = 1.825 \units{ms}$. Success probability of a given number of consecutive passes $n$ decays as $\propto e^{-n/N_\mathrm{pass}}$, consistent with fail events occurring at random. However, probability falls faster for the second check, which may result from drive-induced blinking \cite{brevoord:2024}.

Fitting to the decay rate $N_\mathrm{pass}$ of consecutive passes gives information about the timescale over which the emitter is blinking and/or wandering. For the experiment in Fig.~\ref{fig:readout} (pulse sequence summarized in Fig.~\ref{fig:readout_timing}a), the exponential timescale over which consecutive checks fail is $N_\mathrm{pass}=2.27\pm0.02$ and $N_\mathrm{pass} \tau_\mathrm{cycle} = 4.14 \pm 0.4 \units{ms}$ for the first check of the sequence, and $N_\mathrm{pass}=1.72\pm0.02$ and $N_\mathrm{pass} \tau_\mathrm{cycle} = 3.14 \pm 0.4 \unit{ms}$ for the second check. Finding ways to improve these timescales is an important avenue for further study.

\begin{figure*}[htb!] 
\begin{center}
\includegraphics[width=2.0\columnwidth]{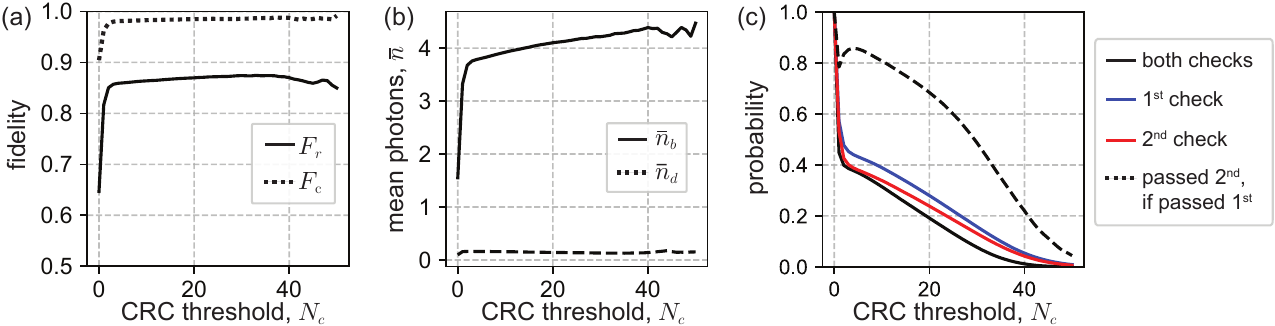}
\caption{
Analysis of the time-tagged dataset shown in Fig.~\ref{fig:readout}, as a function of CRC threshold photon number $N_c$. If both CRCs per cycle each produce a photon number $\ge N_c$, the readout steps in that cycle are kept for analysis. (a) Readout fidelity $F_r$ and conditional fidelity $F_c$ (first readout step and wfm-A, respectively; results from the second readout step and wfm-B look nearly identical). (b) Mean photon number in the bright and dark distributions (first readout step; results from the second step look nearly identical). (c) Probability of passing each CRC, both CRCs, and the probability of the second CRC given a pass of the first. (Data are from wfm-A; results from wfm-B look nearly identical.)
}
\label{fig:readout_vs_N}
\end{center}
\end{figure*}

\subsubsection{Readout fidelity vs power}
\begin{figure}[htb!] 
\begin{center}
\includegraphics[width=1.0\columnwidth]{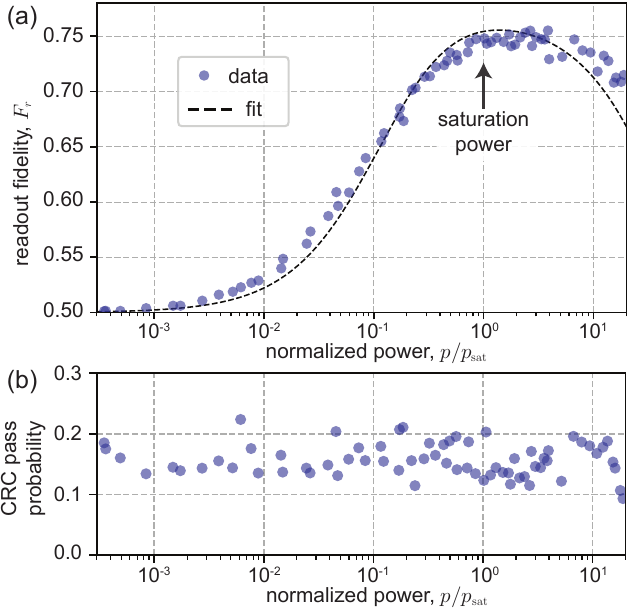}
\caption{
(a) Readout fidelity vs power. Data are fit to Eq.~\ref{eqn:Fr_model}. (b) Fraction of CRCs passed, vs variable power of a readout step placed before the CRC. Both panels use the same dataset as Fig.~\ref{fig:power}.}
\label{fig:fidelity_vs_power}
\end{center}
\end{figure}

Studying the power dependence of readout is important for the optimization of fidelity. Notice in Fig.~\ref{fig:power}b that, while the number of signal photons $\bar{n}_b - \bar{n}_d$ plateaus near saturation, the total number of collected photons continues to increase with power. We attribute this rise to scatter of the excitation laser into the collection path (see Appendix~\ref{sec:master_equation} for details).

This power dependence makes clear that readout fidelity can be optimized by operating near saturation. Below saturation, increasing power simply produces more signal. Above saturation, however, driving with more power does not increase signal and can only cause rising noise. To test this intuition, we plot readout fidelity vs power in Fig.~\ref{fig:fidelity_vs_power}a, measured from the same dataset used to produce Fig.~\ref{fig:power}b. Data are fit to the model in Eq.~\ref{eqn:Fr_model}. This model assumes that emitted photons follow Poissonian statistics; see Appendix~\ref{sec:master_equation} for a derivation.

Note that this fit uses $p_\mathrm{sat} = 313\units{nW}$ and $\eta=0.992\times10^{-3}$ obtained from fits to Fig.~\ref{fig:power}b, and fits for $f_0=0.89\pm0.032$ and $\Lambda=977\pm72$ as free parameters. Fixing $\Lambda=2244$ as a fixed parameter, as expected from the data in Fig.~\ref{fig:intro}, does not produce a fit which closely matches the data in Fig.~\ref{fig:power}b. This discrepancy could be attributed to non-Poissonian statistics of light emitted from the SnV$^-$ (an assumption of Eq.~\ref{eqn:Fr_model}), and/or from detrimental effects due to SnV$^-$ instability (e.g. cycles where the CRC is passed, but a blinking event occurs before or during the readout step). As such, the models in Eqs.~\ref{eqn:emission_model}~and~\ref{eqn:Fr_model} should be taken as simplifications that do not capture every possible effect on readout of the system, and the parameters reported from these fits should be taken as estimates based on these simplified assumptions only.

Finally, to verify the effect shown in Fig.~\ref{fig:dephasing} is truly measurement-induced dephasing, not, for example, drive-induced charging and/or heating, we study the fraction of CRCs passed as a function of weak measurement power. This is done using the same dataset as in Fig.~\ref{fig:power}b, and using a CRC at the end of the cycle after the weak measurement step. Results are show in Fig.~\ref{fig:fidelity_vs_power}b: As weak measurement power is increased, the fraction of passed checks changes little at $0.16 \pm 0.03$, for all data except the highest measured power (at 19 times saturation) in which the fraction of passed checks drops to $0.09$. This suggests that the power-dependent increase in $\Gamma_\phi$ observed in Fig.~\ref{fig:dephasing} is not due to an effect which causes power-dependent blinking and/or spectral diffusion. This finding is consistent with prior demonstration of coherent spin control using an optical Raman drive \cite{debroux:2021}, a technique which cannot work if a detuned laser causes significant error.

\subsubsection{Coherence time}
\begin{figure}[htb!] 
\begin{center}
\includegraphics[width=1.0\columnwidth]{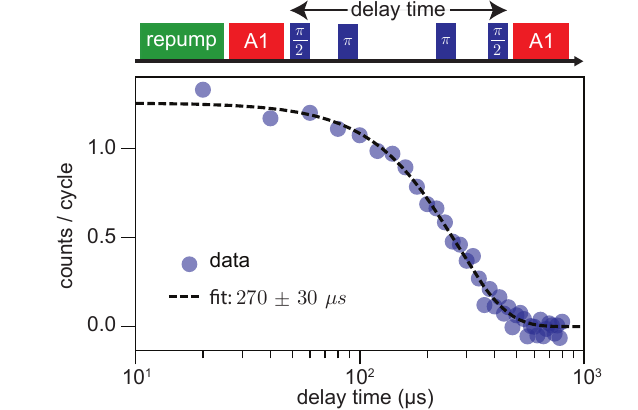}
\caption{
Spin coherence time measured with a CPMG-2 sequence \cite{carr:1954}, measured at $|\Vec{B}|=125 \units{mT}$ oriented at $\zeta=147\degree$. Fitting to $e^{-(t/\tau)^\xi}$ yields $\tau = 270 \pm 30 \units{\mu s}$ with stretch factor $\xi = 1.93 \pm 0.01$.
}
\label{fig:cpmg2}
\end{center}
\end{figure}

The qubit's coherence time is measured to be $T_2^\mathrm{CPMG-2} = 270 \pm 30 \units{\mu s}$ using a CPMG-2 pulse sequence \cite{carr:1954} (Fig.~\ref{fig:cpmg2}) operating at $|\Vec{B}|=125 \units{mT}$ and $\zeta=147\degree$. This measurement is comparable to results in Ref.~\cite{rosenthal:2023}, which operate at $\Lambda\approx80$. For spin-1/2 group-IV qubits, at some magnetic field orientations, dynamical decoupling has a limited effect because the qubit's $g$ factor is too similar to the predominantly spin-1/2 bath \cite{rosenthal:2023}. The measurement in Fig.~\ref{fig:cpmg2} shows that here we avoid this detrimental effect. 

\subsubsection{Quantum jumps}
\begin{figure*}[htb!] 
\begin{center}
\includegraphics[width=2.0\columnwidth]{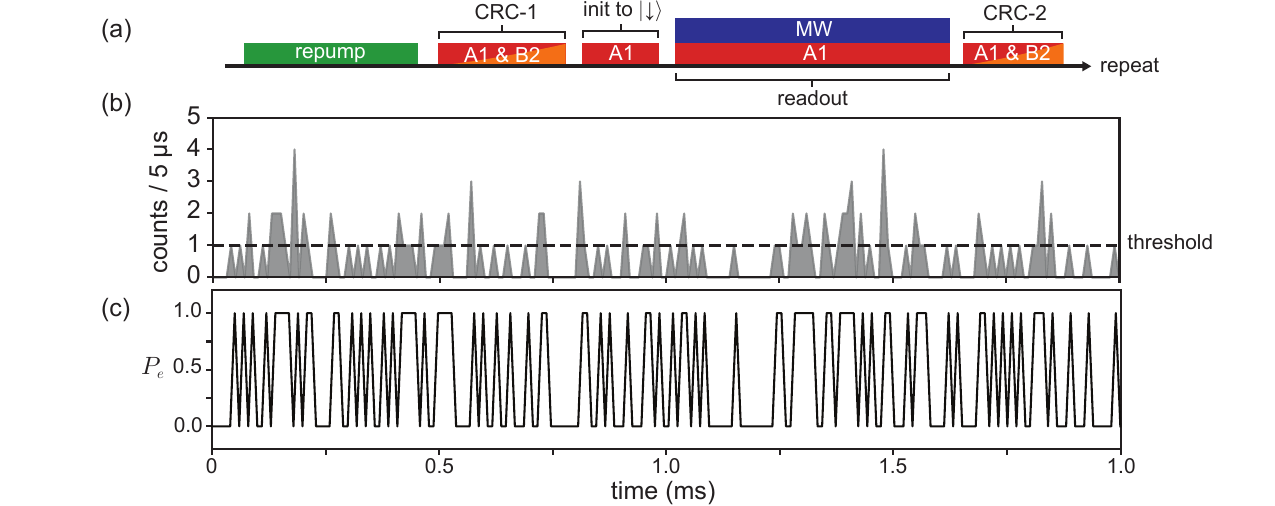}
\caption{
Discrete quantum jumps of the electron spin measured continuously. (a) The measurement sequence consists of a green repump, followed by spin-initialization and continuous readout of the same state in the presence of a weak microwave drive (approximately $1~\units{dBm}$ at the sample). CRCs are implemented before and after the readout step. (b) Time trace of the collected fluorescence counts (in $5~\unit{\mu s}$ bins) while continuous readout of the spin state, showing discrete ``high'' ($N_r \geq1$)  which are real-time jumps of the electron.} (c) The same time trace of the quantum jumps with assigned spin state as 0 or 1 based on the discrimination threshold ($N_{r}$).
\label{fig:quantum_jumps}
\end{center}
\end{figure*}

We take advantage of single-shot readout to observe real-time quantum jumps of the electron spin \cite{robledo:2011}. The protocol, illustrated in Fig.~\ref{fig:quantum_jumps}a, consists of a green repump, then spin polarization, followed by readout on the same state that was polarized in the presence of a weak microwave drive. During this readout step, time-tagged photon detection events are binned into $5\units{\mu s}$ windows and plotted as a time trace in Fig.~\ref{fig:quantum_jumps}b. In the absence of the weak spin drive, we expect zero photons due to high-fidelity initialization of the spin state and, thus, a complete dark time trace of the collected fluorescence. However, the weak spin drive (microwave power approximately $1~\units{dBm}$ at the sample) occasionally kicks the spin state from $\ket{\uparrow}$ to $\ket{\downarrow}$, for example, when initialized using the A1 transition. This results in photons being emitted which can be seen as discrete spikes in the time trace as our readout determines the state in $\ket{\downarrow}$. Using a photon threshold of $N_r\geq1$ (black dashed line), we distinguish between the bright and dark states and denote the electron being in the $\ket{\downarrow}$ ($\ket{\uparrow}$) state. Optical reinitialization also happens naturally under the continuous pump, so the jumping between the spin states (bright and dark) is a mixture of bidirectional mixing (microwave) and one-way reinitialization (bright to dark). To ensure capturing the quantum jumps is not convoluted with charge or spectral hopping, the entire sequence is bracketed with CRCs before and after each measurement shot (using a photon threshold of $N_c=30$ for both CRCs). In this experiment, the spin-flip process from $T_1$ (the order of hertz at the 1.7 K sample temperature \cite{rosenthal:2023,guo:2023}) is much slower than the readout process. For clarity, Fig.~\ref{fig:quantum_jumps}c shows the same time trace where the electronic state is assigned as 0 or 1, based on the discrimination threshold ($N_{r}$).

\subsubsection{Measurement-induced dephasing control experiment}
\begin{figure}[htb!] 
\begin{center}
\includegraphics[width=1.0\columnwidth]{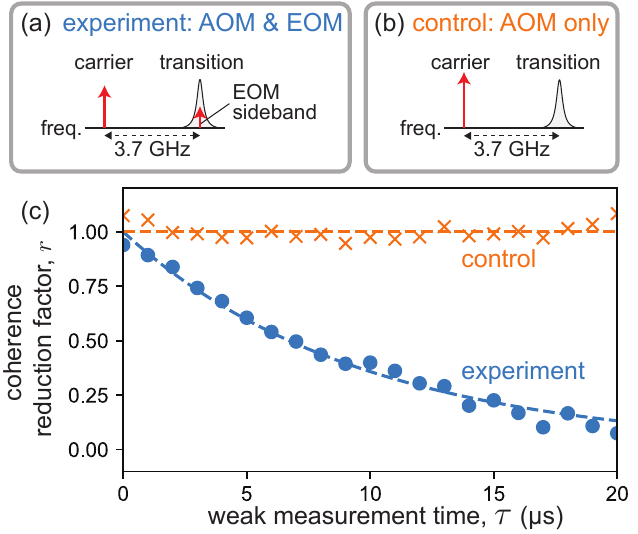}
\caption{
Comparison between the measurement-induced dephasing experiment (as in Fig.~\ref{fig:dephasing}, with pulse sequence in Fig.~\ref{fig:readout_timing}c) to a control. In (a) the experiment, the weak measurement pulse is applied by modulating both the laser amplitude with an acousto-optic modulator (AOM) and modulating an electro-optic modulator (EOM) whose first sideband (here at 3.718 GHz) is resonant with one of the SnV$^-$'s spin-preserving optical transitions. (b) The control sequence is otherwise identical but does \textit{not} modulate the EOM during the weak measurement pulse, so that all laser power remains in the detuned carrier. (c) Measurement-induced dephasing is observed \textit{only} when resonantly driving the transition. Data are shown for a power of 1.3 nW entering the cryostat during the weak measurement pulse, causing a dephasing rate of $\Gamma_\phi/2\pi = 104 \pm 11 \units{kHz}$.
}
\label{fig:mid_control}
\end{center}
\end{figure}

Finally, we do a control experiment to demonstrate again that measurement-induced dephasing is indeed responsible for the dephasing shown in Fig.~\ref{fig:dephasing} rather than any alternative laser-induced process. This experiment is described in Fig.~\ref{fig:mid_control} and consists of comparing dephasing due to a resonant weak measurement pulse to dephasing caused by a control pulse that is off resonant only. No appreciable dephasing is observed during the control pulse, emphasizing that the dephasing observed in Fig.~\ref{fig:dephasing} is indeed caused by measurement.

\subsection{List of data}
\label{sec:operating_conditions}

In Table~\ref{tab:operating_conditions} we provide a summary of data and operating conditions.

\begin{table*}[htb]
\caption{Summary of data and operating conditions. Repump power is typically the order of $100 \units{\mu W}$ and resonant excitation power is typically the order of $1-10 \units{\mu W}$, both specified going into the cryostat. Note that in all data the qubit is driven with the first sideband of the laser created by an electro-optical modulator, not the carrier signal.}
  \begin{center}
    \begin{tabular}{ |p{3.0cm}|p{1.5cm}|p{2cm}|p{2cm}|p{2cm}|p{1.5cm}|p{1.5cm}|p{1.5cm}|p{1.0cm}|  }
         \hline
         Data & Figure & Magnet: \mbox{amplitude $|\Vec{B}|$} & Magnet: \mbox{angle $\zeta$} & Repump duration & Init. duration & Readout duration & Cycle time & CRC? \\
         \hline
         \hline
         Transitions vs $\zeta$ & Fig.~\ref{fig:intro}c & 180 mT & Swept & $1 \units{\mu s}$ & & $1 \units{\mu s}$ & $22 \units{\mu s}$ & No \\
         \hline
         Spin polarization & Fig.~\ref{fig:intro}d & 180 mT & $53\degree$ and $147\degree$ & $100 \units{\mu s}$ & $100 \units{\mu s}$ & & $202 \units{\mu s}$ & No \\
         \hline
         Cyclicity vs $\zeta$ & Fig.~\ref{fig:intro}e & 180 mT & Swept & $100 \units{\mu s}$ & $100 \units{\mu s}$ & & $202 \units{\mu s}$ & No \\
         \hline
         Rabi & Fig.~\ref{fig:intro}f & 125 mT & $147\degree$ & $500 \units{\mu s}$ & $100 \units{\mu s}$ & $100 \units{\mu s}$ & $719 \units{\mu s}$ & No \\
         \hline  
         Rabi vs $\zeta$ & Fig.~\ref{fig:intro}g & 132-183 mT & $147\degree$ & $20-500 \units{\mu s}$ & $5-120 \units{\mu s}$ & $1-60 \units{\mu s}$ & $44-679 \units{\mu s}$ & No \\
         \hline     
         Single-shot readout & Fig.~\ref{fig:readout} & 125 mT & $147\degree$ & $1000 \units{\mu s}$ & $100 \units{\mu s}$ & $50 \units{\mu s}$ & $1825 \units{\mu s}$ & Yes \\
         \hline   
         Emission vs $p/p_\mathrm{sat}$ & Fig.~\ref{fig:power} & 125 mT & $147\degree$ & $500 \units{\mu s}$ & $100 \units{\mu s}$ & $70 \units{\mu s}$ & $1825 \units{\mu s}$ & Yes \\
         \hline  
         Dephasing & Fig.~\ref{fig:dephasing} & 125 mT & $147\degree$ & $500 \units{\mu s}$ & $100 \units{\mu s}$ & $70 \units{\mu s}$ & $1552 \units{\mu s}$ & No \\
         \hline         
         Scattering pattern & Fig.~\ref{fig:scattering}a & 180 mT & $125\degree$ & $10 \units{\mu s}$ & & $10 \units{\mu s}$ & $122 \units{\mu s}$ & No \\
         \hline      
         $F_r$ vs $p/p_\mathrm{sat}$ & Fig.~\ref{fig:fidelity_vs_power} & 125 mT & $147\degree$ & $500 \units{\mu s}$ & $100 \units{\mu s}$ & $70 \units{\mu s}$ & $1825 \units{\mu s}$ & Yes \\
         \hline           
         CRC statistics & Fig.~\ref{fig:crc} & 125 mT & $147\degree$ & $1000 \units{\mu s}$ & $100 \units{\mu s}$ & $50 \units{\mu s}$ & $1825 \units{\mu s}$ & Yes \\
         \hline     
         $F_r$ vs $N_c$ & Fig.~\ref{fig:readout_vs_N} & 125 mT & $147\degree$ & $1000 \units{\mu s}$ & $100 \units{\mu s}$ & $50 \units{\mu s}$ & $1825 \units{\mu s}$ & Yes \\
         \hline              
         CPMG-2 & Fig.~\ref{fig:cpmg2} & 125 mT & $147\degree$ & $500 \units{\mu s}$ & $100 \units{\mu s}$ & $80 \units{\mu s}$ & $3174 \units{\mu s}$ & No \\
         \hline
         Quantum jumps & Fig.~\ref{fig:quantum_jumps} & 125 mT & $147\degree$ & $500 \units{\mu s}$ & $200 \units{\mu s}$ & $1000 \units{\mu s}$ & $2123 \units{\mu s}$ & Yes \\
         \hline 
         Dephasing control & Fig.~\ref{fig:mid_control} & 125 mT & $147\degree$ & $500 \units{\mu s}$ & $100 \units{\mu s}$ & $50 \units{\mu s}$ & $3433 \units{\mu s}$ & No \\
         \hline           
    \end{tabular}
  \label{tab:operating_conditions}
  \end{center}
\end{table*}

\bibliography{main}

\end{document}